\documentclass[11pt,a4paper]{article}

\usepackage{authblk}
\usepackage[utf8]{inputenc}
\usepackage[T1]{fontenc}
\usepackage{amsmath}
\usepackage{amsfonts}
\usepackage{amssymb}
\usepackage{graphicx}
\usepackage{cite}
\usepackage{subcaption}
\usepackage{comment}
\usepackage[table]{xcolor}
\usepackage[pdftex]{hyperref}
\usepackage{tikz}
\newcommand*\circled[1]{\tikz[baseline=(char.base)]{
		\node[shape=circle,draw,inner sep=1.5pt] (char) {#1};}}

\usepackage[left=0.7in, top=0.4in, total={7.0in,10.5in}]{geometry}

\usepackage{acronym}
\newacro{AI}{artificial intelligence}
\newacro{NN}{neural network}
\newacro{DNN}{deep neural network}
\newacro{CNN}{convolutional neural network}
\newacro{RNN}{recurrent neural network}
\newacro{NN}{neural network}
\newacro{ANN}{artificial neural network}
\newacro{SNN}{spiking neural network}
\newacro{ReRAM}{resistive RAM}
\newacro{MTJ}{magnetic tunnel junction}
\newacro{DW}{domain-wall}
\newacro{FM}{ferromagnet}
\newacro{HM}{heavy-metal}
\newacro{AFM}{antiferromagnetic}
\newacro{LIF}{leaky integrate-and-fire}
\newacro{IF}{integrate and fire}
\newacro{MTJ}{magnetic tunnel junction}
\newacro{LTP}{Long-Term Potentiation}
\newacro{LTD}{Long-Term Depression}
\newacro{TL}{top-layer}
\newacro{BL}{bottom-layer}
\newacro{MNIST}{Modiﬁed National Institute of Standards and Technology}
\newacro{PCM}{phase-change memory}
\newacro{eNVM}{emerging non-volatile memory}
\newacro{SRAM}{static random access memory}
\newacro{LLM}{large language model}
\newacro{ReLU}{Rectified Linear Unit}
\newacro{LIF}{leaky-integrate and fire neuron}
\newacro{STDP}{Spike Time Dependent Plasticity}
\newacro{MLC}{multi-level cell}
\newacro{RRAM}{Resistive Random Access Memory}
\newacro{CBRAM}{Conductive Bridge RAM}
\newacro{LRS}{low-resistance state}
\newacro{HRS}{high-resistance state}
\newacro{GMR}{Giant Magnetoresistance}
\newacro{TMR}{tunnel magnetoresistance}
\newacro{STT}{spin-transfer torque}
\newacro{SOT}{spin-orbit torque}
\newacro{LLGS}{ Landau-Lifshitz-Gilbert-Slonczewski}
\newacro{PL}{pinned layer}
\newacro{FL}{free layer}
\newacro{NLP}{natural language processing}
\newacro{CMOS}{Complementary Metal-Oxide-Semiconductor}

\title{Non-Volatile Memory Devices for Neuromorphic Computing}
\author[1]{Debasis Das}
\affil[1]{Department of Electrical \& Electronics Engineering, BITS Pilani, K K Birla Goa Campus, Sancoale, Goa, India.}
\date{}
\begin{document}
	\maketitle
	
	\noindent This chapter explores the critical role of \acfi{eNVM} devices in advancing neuromorphic computing. 
	Inspired by the human brain's remarkable ability to perform complex tasks like pattern recognition and adaptation, neuromorphic computing offers immense potential for implementing complex \acp{NN} in real hardware. 
	Despite ongoing research into the brain's complexities, its fundamental operation relies on neurons (processing units communicating via electrochemical signals) and synapses (junctions between neurons). 
	The modifiable strength of the synapse is key to learning and memory functionality, allowing the brain to adapt and rewire itself. 
	For dense and low-power circuits in neuromorphic computing, these \ac{eNVM} devices are ideal due to their synaptic-like behavior and non-volatile information storage. Property such as conductance modulation of these devices emulate synaptic plasticity which is crucial for learning, and its scalability makes it promising for large-scale neuromorphic architectures. This chapter will delve into the various types of \ac{eNVM} devices such as \acf{ReRAM}, \ac{PCM}, spintronics-based memory devices and their applications in neuromorphic computing, discussing their challenges and opportunities, and highlighting recent advances in their integration within neuromorphic systems.
	
	\section{Introduction}\label{Sec:Intro}
	Since the invention of transistors, the electronics industry has experienced exponential growth, producing increasingly complex and powerful analog and digital devices.  Although device size has shrunk, enabling denser integration and more powerful processing, these devices are significantly less energy-efficient than the human brain when performing complex tasks. For example, Google DeepMind's AlphaGo, which competed against then-champion Lee Sedol in 2016, consumed an estimated $\sim$1 MW of power, while humans performing similar complex tasks use only $\sim$20 W. This vast disparity in energy consumption highlights the need for more efficient computing paradigms.  Inspired by the brain's architecture, neuromorphic computing offers a promising path towards achieving such efficiency. By mimicking the brain's parallel processing and distributed memory, neuromorphic systems have the potential to perform complex tasks with significantly lower power consumption.  However, realizing this potential requires specialized hardware, particularly memory technologies that can emulate the behavior of biological synapses with high fidelity and energy efficiency. 
	
	Although the intricate workings of the human brain remain a subject of ongoing scientific investigation, its fundamental operational principles are understood to be based on the interplay of neurons and synapses. Neurons, the brain's fundamental processing units, communicate with each other through electrochemical signals.  These signals are transmitted across synapses, which are the junctions between neurons.  Crucially, the strength of the synaptic connection between neurons is not fixed; it can be modified based on experience. Synaptic strength's plasticity, or modifiability, is the key to learning and memory.  Changes in synaptic strength allow the brain to adapt to new information, learn new skills, and form memories.  This dynamic rewiring capability enables the brain to constantly adjust its neural circuits in response to its environment and experiences. 
	Inspired by the human brain's remarkable capabilities, the journey of \acp{ANN} began with the aspiration to replicate its learning and information-processing prowess. Early researchers sought to mimic the brain's fundamental building blocks neurons and synapses by creating simplified mathematical models. While rudimentary, these initial attempts laid the groundwork for developing increasingly sophisticated \ac{NN} architectures. The goal was to create systems that could learn and adapt analogously to the brain, paving the way for advancements in fields like pattern recognition, machine learning, and artificial intelligence.  
	\begin{figure}[!b]
		\centering
		\includegraphics[scale=0.7]{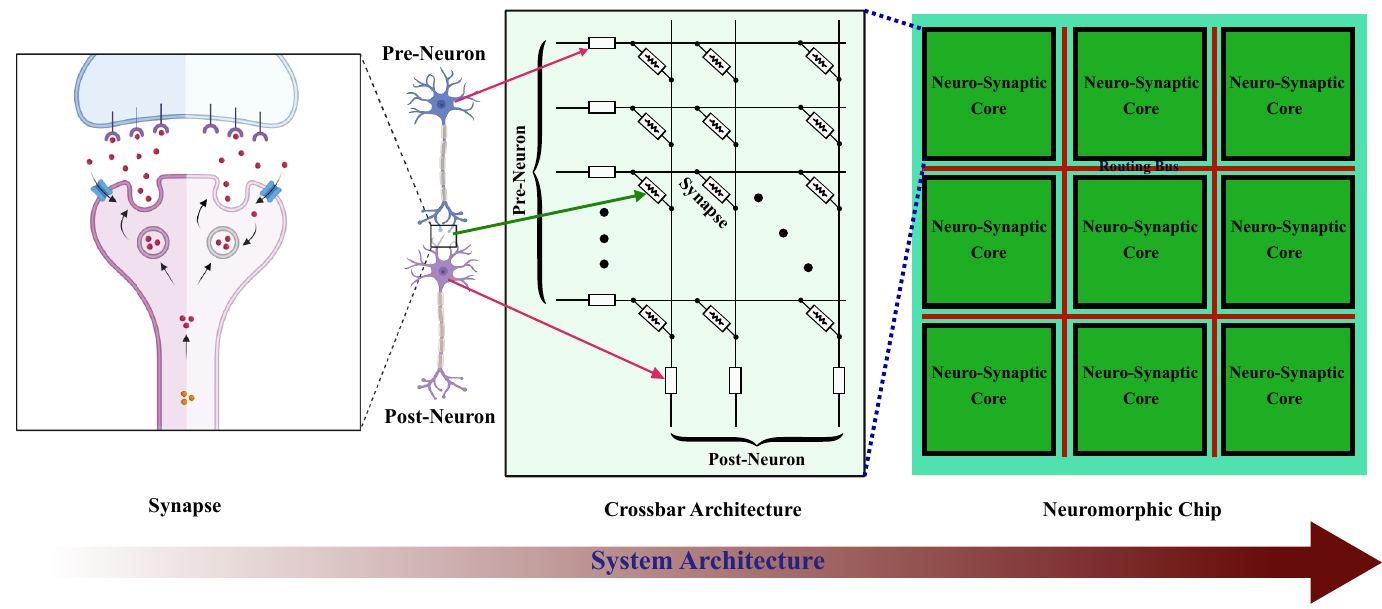}
		\caption{Schematic view of the system architecture of a neuromorphic chip, showing multiple Neuro-Synaptic cores interconnected by a routing bus. Each Neuro-Synaptic core contains a crossbar array connecting pre-neurons to post-neurons via synapses. The biological analogy of neurons and synaptic connections is shown (left).}
		\label{Fig:1}
	\end{figure}
	
	Over the years, \acp{NN} has evolved significantly and is categorized into three generations such as first, second, and third, based on the transfer function of the neurons. In the first generation, \acp{ANN} consisted of McCulloch-Pitts threshold neurons~\cite{maass1997networks}, which generates step function as output. These neurons work in a binary fashion, \emph{i.e.}, if the weighted incoming signal crosses a pre-determined threshold value, it generates a high output. These networks were mostly used for classifying binary patterns. Despite their simple working principle, they give rise to important \ac{NN} models, such as Hopfield nets. Due to its binary nature, these types of \acp{NN} are ideal for boolean computations. Despite the simple working principle of these \acp{NN}, they are not scalable for performing complex pattern recognition tasks. This led to the development of the second generation \acp{NN}, where the step perceptrons were replaced by the neurons that characterize non-linear functions such as sigmoid~\cite{popescu2009multilayer}, tanh~\cite{kim2002fully}, \ac{ReLU}~\cite{li2017convergence}. These non-linear functions enable the \acp{NN} to perform complex tasks, such as image recognition~\cite{teow2000handwritten,liu2003handwritten,phung2007pyramidal}, speech recognition~\cite{chen1996speech,ahad2002speech}, \ac{NLP}~\cite{ma2002natural,goldberg2016primer}, time-series analysis\cite{shiblee2009time}. The patterns are recognized by sending the input data through layers of artificial neurons connected via artificial synapses, leading to a crossbar architecture. One of the milestones for the revolution of \ac{AI}, is due to the invention of the \ac{CNN}, where a hierarchical representation of features is learned automatically from the raw image via convolution operations, leading to a surge of applications in image recognition~\cite{lecun1998gradient,krizhevsky2012imagenet}, computer vision\cite{zhao2017object, ren2016single}, and other related fields. Although second-generation \acp{NN} can perform more complex tasks than its first-generation counterpart and achieve near-human accuracy, this comes with a cost. Apart from the fully connected layers, the convolution layers needed to extract various features of the input images lead to a much deeper \ac{NN} architecture, famously known as \acp{DNN}. Due to these deep layers of the kernels, and neurons, these \acp{NN} consume a significant amount of energy during training as well as inference. With the ever-growing demand and the complexity of such networks, energy consumption has increased on an exponential scale, which is of significant concern, particularly for resource-constrained devices and large-scale deployments. \Acp{SNN} have emerged as a promising alternative to address the energy consumption challenges. Inspired by the biological neuron's event-driven communication via electrical spikes, \acp{SNN}, often considered a third generation of neural networks, offers a fundamentally different approach to computation with sparse, binary "spikes" rather than continuous activations, compared to its previous counterparts. The event-driven nature of spiking neurons means computations are only performed when spikes occur, resulting in sparse activity and increased energy efficiency. Furthermore, the temporal dynamics of SNNs, where information is encoded in the timing of spikes, offer a more biologically plausible and potentially more powerful computational paradigm. This inherent efficiency and biological realism make SNNs a strong candidate for energy-constrained applications and the development of more sustainable AI.
	
	Neuroscience indicates that the human brain's impressive analytical power is a result of its vast network, encompassing billions of neurons linked through trillions of synaptic connections. To emulate the brain's vast interconnectivity in \acp{ANN}, neuro-synaptic connections are commonly implemented using a crossbar architecture, regardless of the network generation. One such architecture is shown in Fig.~\ref{Fig:1}, where a set of pre-neurons from one layer are connected to the set of post-neurons of the subsequent layer via synapses. These crossbars are crucial in the design of neuromorphic chips~\cite{davies2018loihi,cassidy202411}, forming the core building blocks known as neuro-synaptic cores that mimic the brain's structure. These cores are then interconnected via a routing bus, which enables communication and exchange of information between different parts of the chip, as shown in Fig.~\ref{Fig:1}. This distributed architecture allows the chip to perform cognitive tasks by processing information in a parallel and event-driven manner, similar to the brain. Regardless of the neuronal transfer functions such as non-linear activation functions or spiking, the training of the neural networks involves tweaking the synaptic weights to establish proper connections between neurons from subsequent layers. Thus, for realizing the architectures in hardware, memory devices play a crucial role in storing and efficiently updating these synaptic weights. Deployments of such hardware into the real world, especially in resource-constraint environments, demand the devices emulating the synaptic behavior, to be non-volatile, area, and energy efficient. Thus, in this chapter, we explore the various \ac{eNVM} devices spanning from \acf{PCM}, \acf{ReRAM}, and spintronics devices and discuss how they can be utilized to emulate neuro-synaptic behavior. 
	
	\section{Understanding the neuro-synaptic behavior}\label{Sec:2} 
	Apart from the non-volatile nature, and smaller size, \ac{eNVM} devices are capable of mimicking the neuro-synaptic behavior of the biological neurons and synapses.  
	In Sec.\ref{Sec:Intro}, we discussed the \acp{NN} are categorized based on the neuronal transfer function. This section provides further details on the transfer functions in a generic perspective, along with the synaptic behavior for training the \acp{NN}. 
	\begin{figure}[h!]
		\centering
		\includegraphics[scale=0.45]{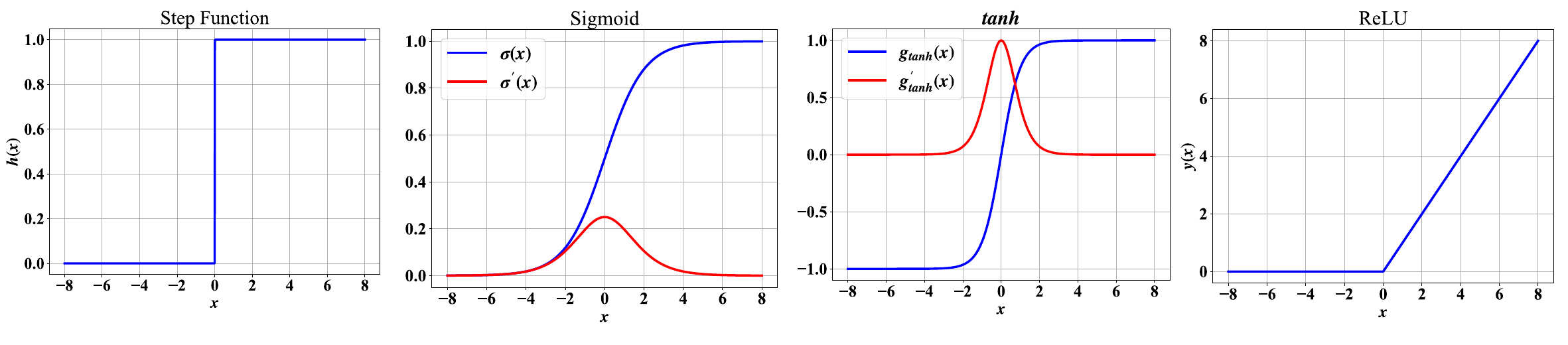}

		\caption{Plot of neuronal transfer functions. (a) Step-function (first generation). Second-generation transfer functions) (b) Sigmoid, (c) Hyperbolic tangent (tanh), and (d) \acf{ReLU} function.}
		\label{Fig:2}
	\end{figure}
	\subsection{Neuronal dynamics}
	We start with the first generation neural transfer function as illustrated in Fig.\ref{Fig:2}a, which is a step function, that produces a binary output based on the input $x$.  The output is high if the input crosses a threshold (in this case, $x=0$), and low otherwise. While the step function can handle simple pattern recognition, its abrupt, non-differentiable nature prevents it from effectively learning complex patterns using gradient-based techniques, such as backpropagation. On the contrary, second generation \acp{NN}, overcome this limitation, thanks to their gradual and smooth transitions. For example, the sigmoid function's $\left(\sigma(x)=\frac{1}{1+e^{-x}}\right)$ (shown in Fig.~\ref{Fig:2}b) smooth, probabilistic nature (with output values ranging between 0 and 1) enables the network to learn more complex patterns and classify data with greater confidence. In addition to Sigmoid, another widely used activation function is the hyperbolic tangent, denoted as $tanh$ function $\left(g_{tanh}(x)=\frac{e^x-e^{-x}}{e^x+e^{-x}}\right)$, as shown in Fig.~\ref{Fig:2}c. Although both activation functions have similar $S$-shaped curves, but there are a few key differences that make $tanh$ a better choice. For example, the output of the $tanh$ ranges between -1 and +1, with a zero-centered, whereas the sigmoid's output ranges between 0 and 1, always, positive. Thus, the gradient during backpropagation, tends to be more balanced for $tanh$, whereas the positive output of sigmoid, can lead to a bias in the gradients. Also, $tanh$ has a steeper gradient around zero, potentially leading to a faster convergence during training. However, both these functions suffer from vanishing gradient problems, where the derivative approaches zero or ``vanishes'' at higher positive and negative values of $x$. As a result, the constituent layers of the \ac{NN} receive little to no information for weight updates during backpropagation, hindering the overall training process. On the other hand, \ac{ReLU} (shown in Fig.\ref{Fig:2}d) offers a significant advantage over the sigmoid and $tanh$, due to its ability to mitigate the vanishing gradient problem. With a consistent gradient of 1, for the positive inputs, ReLU ensures that gradients don't diminish as they propagate through multiple layers, helping to train the \ac{NN} effectively. The simplicity of the ReLU activation function contributes to its computational efficiency, resulting in faster training and inference times. Leveraging these activation functions, second-generation \acp{NN} made substantial progress in addressing sophisticated pattern analysis, as exemplified by architectures like AlexNet~\cite{krizhevsky2012imagenet}. 
	
\begin{figure}[!hbt]
		\centering
		\includegraphics[scale=0.4]{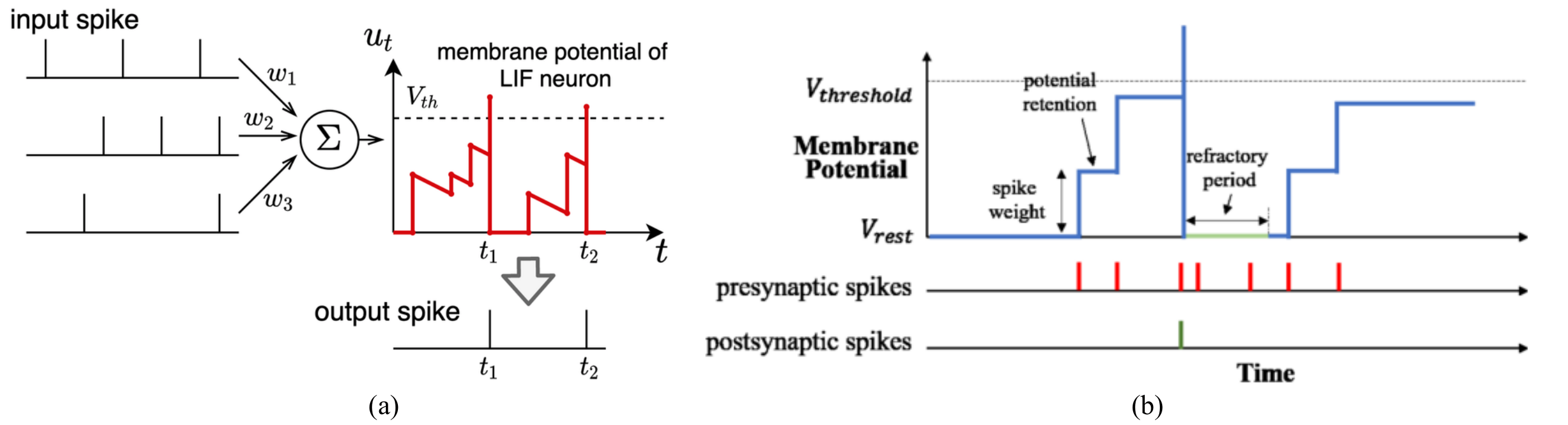}
		
		\caption{(a) LIF neuron model is showing the pre-synaptic pulses get modified according to the weights and summed up, and eventually changes the membrane potential (red). Output spike is generated while the membrane potential crosses the threshold value, $V_{th}$. (b) Membrane potentials and spikes are shown for the IF model. The images are taken from Ref.~\cite{kamata2022fully}, and \cite{zhou2021improved}, respectively.}
		\label{Fig:3}
	\end{figure}

	Even with the impressive cognitive abilities they possess, second-generation neural networks cannot match the human brain's energy efficiency due to their continuous operation across the entire network. In contrast, \acfp{SNN}, often called the third generation neural networks, aim to address this energy efficiency challenge by employing event-driven communication using discrete ``spikes'', resulting in sparse neural activity. \ac{SNN} neurons process information by receiving ``spikes''-brief electrical pulses from pre-synaptic neurons. These incoming signals are then modulated by the synaptic weights connecting the pre and post-synaptic neurons. These modulated signals change the membrane potential of the neurons, a phenomenon, known as `Integration'. There exist different neuron models, such as \ac{IF}, and \ac{LIF} neurons. The \ac{LIF} neuron's membrane potential decays when there are no pre-synaptic spikes, a phenomenon, known as `leak' (as shown in Figure~\ref{Fig:3}a), whereas for the \ac{IF} neuron, the membrane potential remains constant (as shown in Figure~\ref{Fig:3}b), under such conditions. Through these repetitive processes, if the membrane potential crosses a pre-determined threshold value, it emits a spike, as shown in Figure~\ref{Fig:3}a and \ref{Fig:3}b. Afterward, the neuron remains inactive for a certain period, referred to as \textit{refractory period}, \emph{i.e.} incoming spikes do not affect the neuron.
	
	\subsection{Synaptic behavior}
	Synapses are the fundamental connections between the neurons, crucial for information processing and learning in neural networks. Each synapse has a weight that modulates the strength of the signal, influencing the postsynaptic neuron's response. Biological neural networks exhibit synaptic plasticity, a process where synaptic weights are modified during the learning phase. After learning is complete, stable synaptic weights are essential for consistent cognitive performance. Therefore, devices mimicking artificial synapses must exhibit plasticity, and non-volatility to retain the weights after training, for the \emph{inference}. Before, discussing the details about the functionality of the \ac{eNVM} devices as artificial neurons and synapses, we need to know the process of modifying the synaptic weight during learning, which is categorized as supervised and unsupervised learning.
	
	\subsubsection{Supervised learning}
	Neural networks are trained for specific tasks, such as computer vision, and speech recognition, with the corresponding datasets (such as \ac{MNIST} hand-written digit dataset, CIFAR 10, \emph{etc}). The goal of neural network training is to modify the synaptic weights so that the network can accurately classify new, unseen data similar to the training data. 	
	Supervised learning, a catogory of of machine learning that involves training algorithms on labeled datasets, which provide both input characteristics and their matching target output values.	
	The training of the neural network is done using the backpropagation~\cite{nielsen2015,lillicrap2020backpropagation} method, where, the gradient of the loss functions (calculated from the predicted and labeled output) is used to adjust the weights and biases of the synapses and neurons, respectively. Given input propagates through the network, and produces output, a method known as \emph{forward pass}. The error is calculated by comparing the predicted label with the true label, \emph{i.e.} calculating the loss function. The backpropagation algorithm then performs a backward pass, propagating the error signal back through the network, layer by layer. During this backward pass, the gradient of the loss function is computed with respect to each weight. The sign and magnitude of the computed gradient denote the direction and change needed in the weights to minimize the loss function. This process is done iteratively unless the loss function is minimized. The term `supervised learning' arises from the use of labeled data, which offers explicit guidance to the algorithm during training, allowing it to discern the inherent connections between inputs and outputs.
	Despite the remarkable achievements of supervised learning, in the field of computer vision, \ac{NLP}, it still doesn't fully mirror the learning processes observed in biological brains. This highlights the need for an alternative learning mechanism that resembles the biological learning mechanism and potentially introduces more energy efficiency and adaptability. 
	
	\subsubsection{Unsupervised learning}
	Unsupervised learning, unlike its supervised counterpart, does not rely on labeled data for training the neural network, instead, it leverages localized learning rules to detect patterns and structure within the data itself. To truly, capture the essence of the biological neural network, in terms of energy efficiency, the mode of communication between the neurons can be implemented using \emph{spikes}, a discrete event produced by the neurons, that makes the network sparse and enhances the energy efficiency by reducing the computational complexities. Most widely known unsupervised learning algorithm in spike-based systems is \ac{STDP}. It adjusts synaptic weights by strengthening or weakening connections based on the relative timing of pre- and post-synaptic spikes, inspired by the rat's hippocampal glutamate synapse. The increase of the synaptic weight is known as \ac{LTP}, whereas \ac{LTD} refers to the decrease in weight.  
	\ac{LTP}, which strengthens the connection weight, occurs when the post-synaptic firing occurs before the pre-synaptic firing, while \ac{LTD}, which weakens the connection, happens when the post-synaptic neuron fires after the pre-synaptic neuron. Also, the change in synaptic weight is relative to the temporal difference between the action potentials, a relationship captured by the following mathematical model~\cite{bi2001synaptic,cai2014neuronal}:
	\begin{equation}
		\begin{split}
			\delta(\Delta t) &= A^{+}exp\left(-\Delta t/\tau^+\right)~~~\mathrm{if} ~\Delta t > 0,\\
			&= A^{-}exp\left(\Delta t/\tau^-\right)~~~\mathrm{if} ~\Delta t < 0,
		\end{split} 
	\end{equation}
	where, $A^+$, $A^-$, $\tau^+$, $\tau^-$, are the amplitude and the time constants for the weight modification. Here, $\Delta t$ ignifies the time interval between a pre-synaptic spike and a post-synaptic spike. 
	
	\section{Implementing neuromorphic hardware using \ac{eNVM} devices}
	Due to its non-volatile nature, and the inherent device physics, \ac{eNVM} devices can mimic the neuro-synaptic behavior. Their compatibility with the traditional \ac{CMOS} technology makes it suitable to implement neuromorphic hardware, by reducing area footprint. As discussed, trained neural network weights require non-volatile storage, which can be readily implemented using eNVM devices in a crossbar architecture. Moreover, the ability to program and erase the information in these \ac{eNVM} devices allows them to be utilized for unsupervised learning, where the weights are determined through online training.    
	
	In this section, we first discuss the device physics of various \ac{eNVM} technologies, such as \ac{PCM}, resistive \ac{ReRAM}, and spintronics devices, and how these characteristics can be manipulated to mimic the neuro-synaptic behavior. As in the previous sections, it is well understood that \acp{SNN} offers much more energy efficiency; thus our focus would be on discussing the implementation of the \ac{SNN} hardware using \ac{eNVM} devices. We will also discuss the crossbar architecture of these devices, leading to a system-architecture perspective.
	
	\subsection{Phase change memory devices}
	\subsubsection{Device Physics}
\begin{figure}[!h]
		\centering
		\includegraphics[scale=0.2]{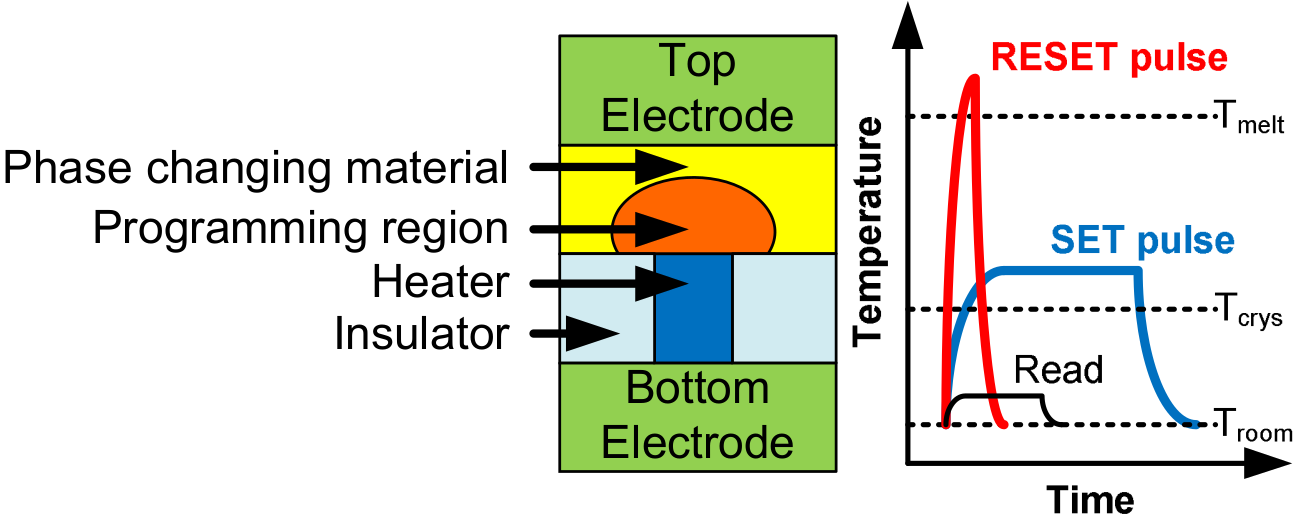}
		\caption{Schematic device structure of the \ac{PCM}, with \ac{PCM} material inserted between the top and bottom electrodes. Corresponding RESET, SET, and Read programming pulses are shown. The images are taken from Ref~\cite{jeyasingh2012phase}.}
		\label{Fig:PCM}
	\end{figure}

	\Acf{PCM} devices rely on the reversible change of phase between the amorphous and crystalline state of the chalcogenide glass material. The amorphous state possesses high electrical resistivity, whereas the crystalline state offers low electrical resistance, with a ratio as high as $\sim10^3$~\cite{stern2021uncovering}. These two distinct states are utilized to use \ac{PCM} devices as memory for storing information. In a typical \ac{PCM} device, phase change material is fabricated on top of a narrow heater enclosed in an insulator. This combined structure is then sandwiched between the top and bottom electrodes, as shown in Figure~\ref{Fig:PCM}. The transformation between the amorphous and crystalline states is achieved through controlled heating and cooling cycles by applying necessary electrical pulses between the electrodes. To switch a PCM cell from its crystalline to amorphous state, a short, high-amplitude electrical current pulse is injected, a process known as the RESET operation. Intense Joule heating, resulting from the application of a short pulse, raises the PCM material's temperature beyond its melting point. Following the pulse termination, the rapid cooling, forces the material to solidify in a disordered amorphous state, resulting in a high resistance. To switch the \ac{PCM} cell from the amorphous to the crystalline state, a low amplitude but a longer current pulse is applied. The lower amplitude of the current pulse results in moderate Joule heating, which increases the temperature beyond the crystalline temperature yet keeps it below the melting point. This controlled heating allows atoms to rearrange themselves into an ordered crystalline lattice. After the pulse is removed, the material cools down more slowly and this gradual cooling allows the formation of a crystalline phase, exhibiting lower resistance. To read the memory state, an electrical current pulse with a magnitude significantly lower than the SET pulse is applied to avoid disturbing the current state. By comparing the obtained voltage against a reference, the read state can be determined.
	
	\subsection{\ac{PCM} as Spiking Neurons}
\begin{figure}[!h]
		\centering
		\includegraphics[scale=0.8]{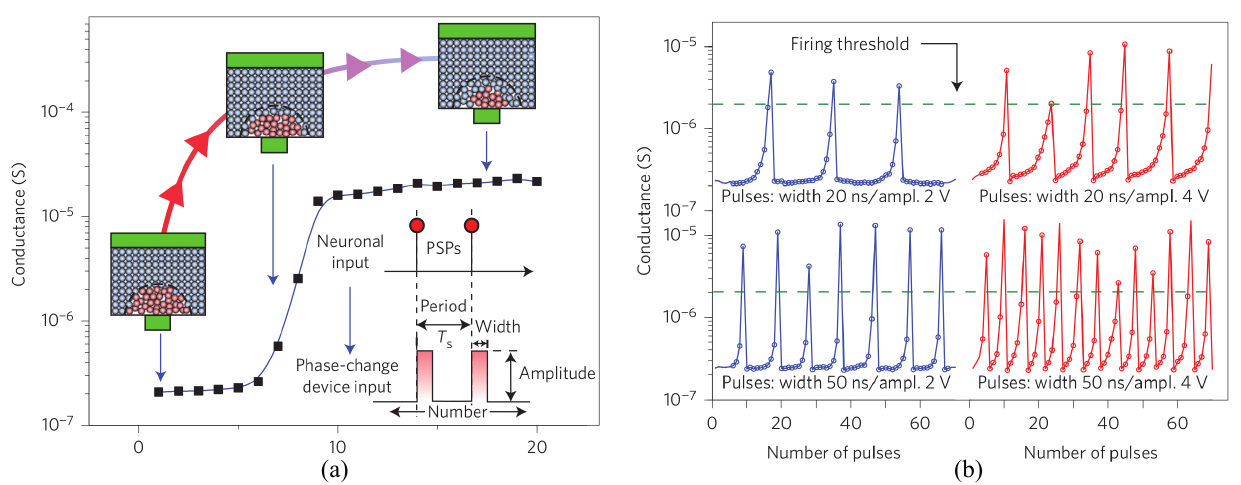}
		\caption{(a) Variation of conductance of a \ac{PCM} device with the crystallizing pulse numbers, showing a gradual change of phase from amorphous to crystalline. (b) Integrate-and-fire dynamics of the device for various pulse widths and amplitudes, with a threshold of 2~$\mu$S. The images are adapted from Ref.~\cite{tuma2016stochastic}. }
		\label{Fig:PCM_Neuron}
	\end{figure}

	We have discussed how \ac{PCM} device operates to store information based on the different conductance values in its amorphous and crystalline phase. We have also discussed how a gradual change from the amorphous to crystalline phase is obtained using comparatively lower magnitude SET pulses. This behavior of the \ac{PCM} device along with high conductance at its crystalline phase can be exploited to mimic the functionality of the spiking neurons. In its initial, or resting, state, a substantial amorphous region covers the bottom electrode, effectively isolating the crystalline phase with a high-resistance barrier. Applying consecutive current pulses to the device, similar to SET pulses, emulates the pre-synaptic pulses received by biological neurons. These pulses can be thought of as crystallization pulses. Tuma \emph{et.al.}~\cite{tuma2016stochastic} experimentally demonstrated that these applied pulses gradually alter the PCM material's phase, resulting in a gradual reduction of the thickness of the amorphous region. The thickness of the amorphous region could be interpreted as the membrane potential of the biological neuron. By repeatedly applying these crystallization pulses, conductance increases as shown in Figure \ref{Fig:PCM_Neuron}a, and beyond a certain threshold value, \ac{PCM} changed to crystalline phase and emits a spike~\cite{tuma2016stochastic}, mimicking \emph{\acf{IF}} neuron. After emitting the spike, biological neurons go through a reset mechanism, by resetting its membrane potential. The spiking behavior is shown in Figure~\ref{Fig:PCM_Neuron}(b). As discussed, the RESET pulse can be used to reset the phase of the \ac{PCM} from crystalline to amorphous. Cobley \emph{et.al.}~\cite{cobley2018self} integrated \ac{PCM} neuron cell into the SPICE model to demonstrate a self-reset functionality, which automatically returns the cell to its initial state after a spike is emitted by using a feedback loop to quickly melt and quench the PCM material.
	
	\subsection{\ac{PCM} as Synapse}
	\begin{figure}[!b]
		\centering
		\includegraphics[scale=0.6]{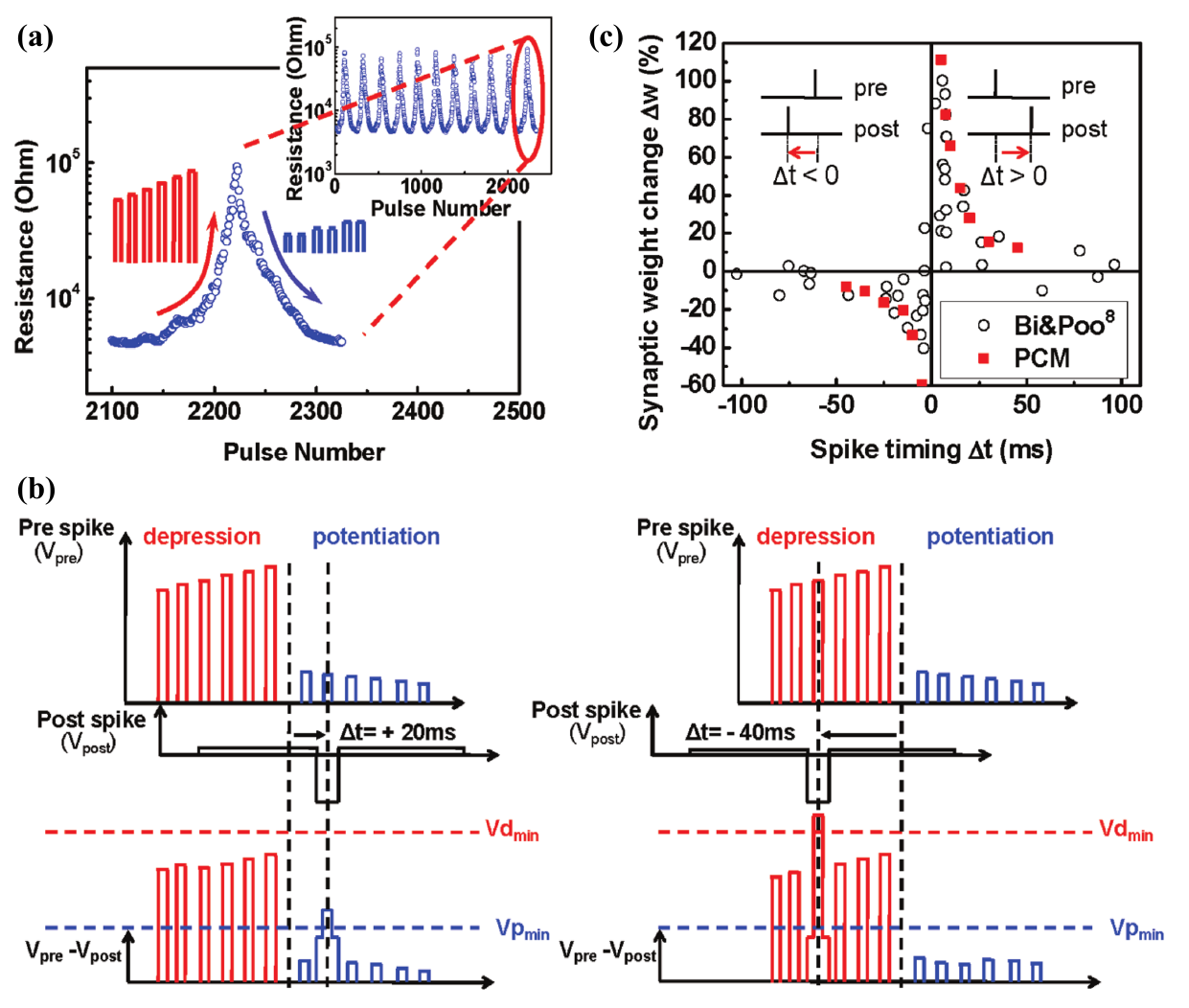}
		\caption{(a) Cell resistance undergoes a gradual RESET process facilitated by pulses of increasing amplitude (red), whereas cell resistance is gradually SET through the application of a staircase-like pulse. Multiple intermediate resistance states are visible encompassing the entire resistance window. (b) Pre- and post-synaptic programming voltage pulses at $\Delta t$=20 ms (left) and $\Delta t$=-40 ms (right) show potentiation and depression. (c) Synaptic weight change is plotted with the relative timing between pre- and post-synaptic spikes and corresponds comparison of the biological synapse. Images are taken from Ref~\cite{kuzum2012nanoelectronic}.}
		\label{Fig:PCM_Synapse}
	\end{figure}
	Unlike spiking neurons, synapses exhibit multiple memory states, allowing the network to learn complex patterns and store varying degrees of synaptic strength, which is essential for information processing and adaptation. Typically, 2-3 orders of magnitude difference between its SET and RESET states in the \ac{PCM} devices, opens up an opportunity to accommodate more intermediate states for the \ac{MLC} storage, instead of the traditional binary memory state. Kuzum \emph{et.al.}~\cite{kuzum2012nanoelectronic} demonstrated the fabricated \ac{PCM} cell can achieve multiple intermediate states between its highest and lowest resistance values. Figure \ref{Fig:PCM_Synapse}(a) shows such multiple intermediate resistance states. Their laboratory-fabricated device exhibited an endurance of $\sim 10^7$ cycles, proving the reliability of such \ac{PCM} based synapse in the neuromorphic system. They also demonstrated how gradually increasing amplitude voltage pulses can be used as depression pulses, whereas stair-case-like pulses are used as potentiation pulses. Based on the timing difference between the pre and post-synaptic spikes \emph{i.e.} $\Delta t=t_{pre}-t_{post}$, the weights of the synapse can be increased or decreased, thus emulating potentiation or depression behavior of the biological synapse as shown in Figure~\ref{Fig:PCM_Synapse}(b). Their result matches well with the synaptic weight modulation of the actual biological synapse~\cite{bi1998synaptic} as shown in Figure~\ref{Fig:PCM_Synapse}(c). Using a 10 ns voltage pulse width, they observed an average energy consumption of approximately 50 pJ for RESET operations and only about 0.675 pJ for SET operations, highlighting PCM devices as promising candidates for high-speed, low-energy neuromorphic computing with online unsupervised learning. For real-world applications, a system must not only learn specific patterns but also demonstrate the ability to forget previous patterns and learn new ones in dynamically changing environments. Therefore, synapses built with eNVM devices should be capable of reconfiguring synaptic weights through learning new patterns while interacting with stimuli from continuously changing, real-world environments. 
	\begin{figure}[!h]
		\centering
		\includegraphics[scale=0.2]{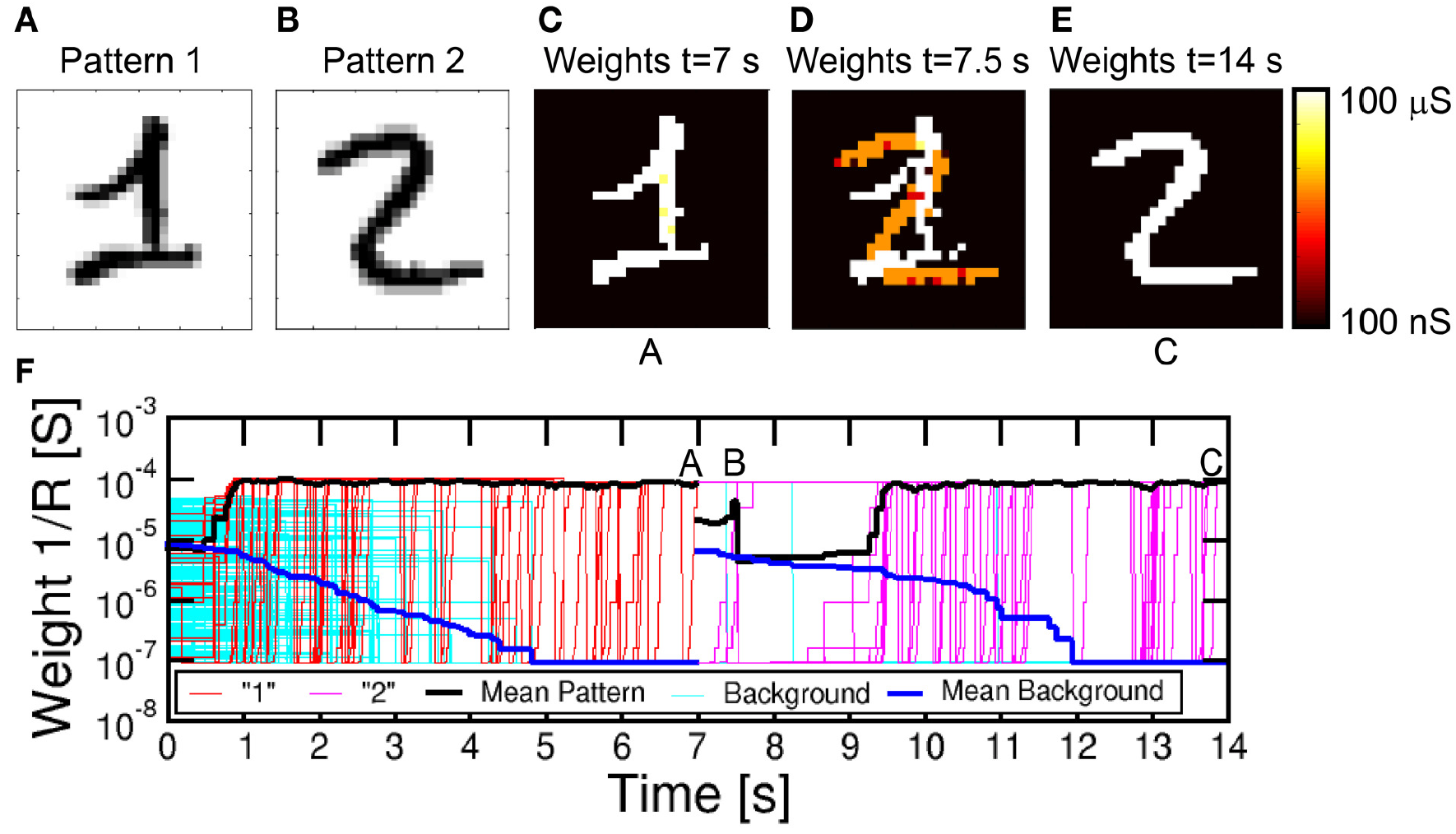}
		\caption{(A) Pattern ``1'' with noise, followed by (B) pattern ``2'' with noise. (C) Learning of pattern ``1'' after 7 seconds. (D) Transition state at 7.5 seconds, showing forgetting of ``1'' and learning of ``2''. (E) Learning of pattern ``2'' at 14 seconds. (F) Temporal evolution of synaptic weights, demonstrating learning of ``1'' and subsequent update to ``2''. Figures are adopted from Ref.~\cite{ambrogio2016unsupervised}.}
		\label{Fig:PCM_Synapse_dynamic}
	\end{figure}
	Ambrogio \emph{et.al.}~\cite{ambrogio2016unsupervised} demonstrated one such event where $N$ pre-synaptic neurons are presented with handwritten digits of `1' for 7 seconds, and then another digit `2', is presented. Figure \ref{Fig:PCM_Synapse_dynamic}(C)-(E) demonstrates how the weights of $N\times M$ synapses are dynamically changing with time, where, $M$ is the number of post-synaptic neurons. The demonstrated endurance, speed, and low-power characteristics of PCM devices, along with their capacity for online learning and pattern adaptation, position them as a promising technology for future neuromorphic hardware. 
	
	\section{Resistive memory device}
	\subsection{Device Physics}
\begin{figure}[!h]
		\centering
		\includegraphics[scale=0.8]{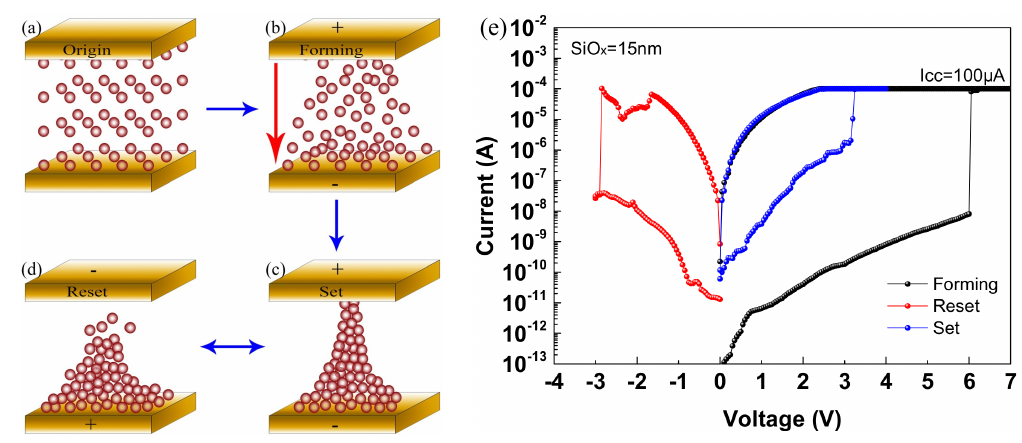}
		\caption{(a)-(d)Schematic diagram showing different conduction mechanism by the formation and rupture of conducting filament. (e) Current vs volatge characteristics demonstrating formation, SET and RESET process. Figures (a)-(d) are adopted from Ref\cite{liu2019resistive} and (e) is adopted from Ref\cite{yen2019all}.}
		\label{Fig:RRAM_IV}
	\end{figure}

	Building upon the promising capabilities of \ac{PCM}, other resistive memory devices also emerge as strong candidates for neuromorphic applications. Recent advancements in materials and electrical switching mechanisms have significantly improved ReRAM performance. Similar to \ac{PCM}, this resistive device operates by modifying the resistance of a material layer confined between its top and bottom metalic contacts. If the sandwiched material is oxide, then we refer it as \ac{RRAM}, whereas if the sandwiched material is a conductive element, we refer it as \ac{CBRAM}. The resistive switching behavior of \ac{RRAM} is governed by the formation of oxygen vacancies and the migration of oxygen ions. Anion-based RRAM usually necessitates either an oxygen-ion active electrode or an oxygen-deficient film to function as an oxygen-ion reservoir during the resistive switching process. This type of are usually switching is observed in oxide-based materials, including metal-oxide-based and silicon-oxide-based materials. The switching process in metal oxide-based RRAM involves transitioning between a \ac{HRS} and a \ac{LRS} through the controlled formation and rupture of conductive filaments~\cite{ielmini2012evidence,yu2013hfox,chu2013charge}. In its pristine state, the metal oxide layer has few or no conductive filaments, resulting in high resistance. SET process, \emph{i.e.} transition from \ac{HRS} to \ac{LRS} is achieved by applying a voltage pulse across the metal oxide layer. In an experimental method, Liu \emph{et.al.} demonstrated that a large number of defects due to oxygen vacancies in $\mathrm{HfO_2}$ film distributed randomly at zero bias voltage, depicting \ac{HRS} (Figure \ref{Fig:RRAM_IV}(a)). When a positive voltage was applied, the traps were filled by the injected electrons, and as the voltage increases, the conducting filament started to grow (Figure \ref{Fig:RRAM_IV}(b)). As the conducting filament starts to form the conductivity increases and so does the current in the \ac{RRAM}, as shown in Figure \ref{Fig:RRAM_IV}(e).
	When the conducting filament forms completely, thus creates a fully conducting path (Figure \ref{Fig:RRAM_IV}(c)) between the two electrodes, which results in a sudden jump in current. To prevent any damage at the device level, a compliance current is used to limit the current ($\mathrm{I_{CC}}$) flowing through the device. For the RESET, we reduce the voltage amplitude and eventually apply a reverse voltage across the electrodes. This reverse voltage initiates the recombination of oxygen ions and vacancies, thereby disrupting the conductive filament. When the conducting filament gets completely ruptured this results in an instant RESET from \ac{LRS} to \ac{HRS}, as shown in Figure \ref{Fig:RRAM_IV}(d) and (e). 
	
	\subsection{RRAM as Spiking Neurons}
\begin{figure}[!h]
		\centering
		\includegraphics[scale=0.8]{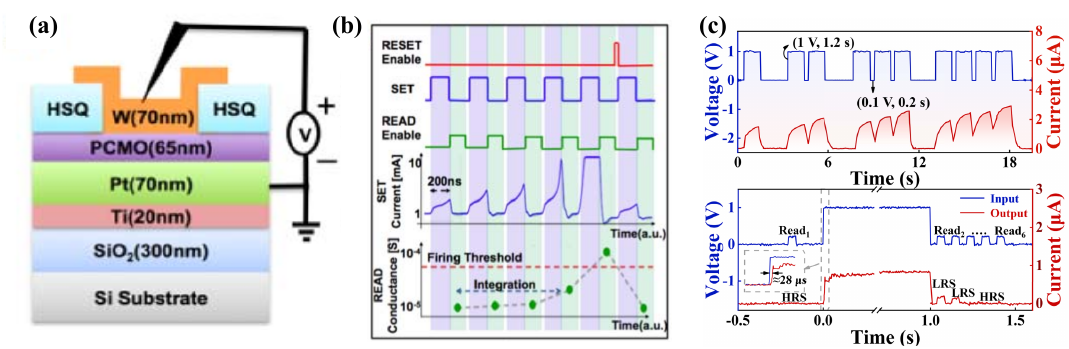}
		\caption{(a) Device schematic of the PCMO-based RRAM. (b) Applied SET, READ, and RESET pulse across the device, and corresponding currents show the integration and firing events for the PCMO-based RRAM. (c) The current response of a stacked memristive device corresponding to the applied voltage pulses. Figures (a) and (b) are adapted from \cite{lashkare2018pcmo}, whereas Figure (c) is adapted from~\cite{bian2022stacked}.}
		\label{Fig:RRAM_Neuron}
	\end{figure}

	In the I-V characteristics of the \ac{RRAM}, we have noticed conductance increases with the application of the voltages and at some certain voltage, the conductance changes sharply. By leveraging these attributes of \ac{RRAM} devices, the behavior of biological neurons can be emulated. Lashkare \emph{et.al.}~\cite{lashkare2018pcmo} demonstrated $\mathrm{Pr_{0.7}Ca_{0.3}MnO_3}$ (PCMO) \ac{RRAM} can be utilized to mimic the \ac{IF} functionality. They initialized the device state to RESET mode, exhibiting lower conductance. Instead of applying a long SET pulse to SET the device, they applied short pulses so that the current rises slowly, but should avoid sharp shoot up. A READ pulse, with a much lower voltage than memory threshold switching, was also applied after each short pulse to measure the conductance. With the repeated application of such short pulses, conductance starts to increase, imitating the integrate functionality of the neuron. Once it reaches a particular value, conductance shoots up in a single pulse, imitating fire functionality. Upon exceeding the threshold of 50 $\mu S$, the device is reset by applying a RESET pulse. The schematic of the device structure is shown in Figure \ref{Fig:RRAM_Neuron}(a), and the current response of the applied SET pulses are shown along with READ and RESET pulse in Figure \ref{Fig:RRAM_Neuron}(b). They used this artificial spiking neuron in a \ac{SNN} architecture to identify Fisher's Iris flower dataset, which shows an accuracy of 95.33\%, compared to an ideal \ac{LIF} neuron's 96\% accuracy. Bian \emph{et.al.}~\cite{bian2022stacked} incorporated the leaky functionality along with the integrate, fire, and reset in a stacked memristive device, thus emulating \ac{LIF} functionality of the biological neuron.  
	
	\subsection{RRAM as Synapse}
\begin{figure}[!b]
		\centering
		\includegraphics[scale=0.22]{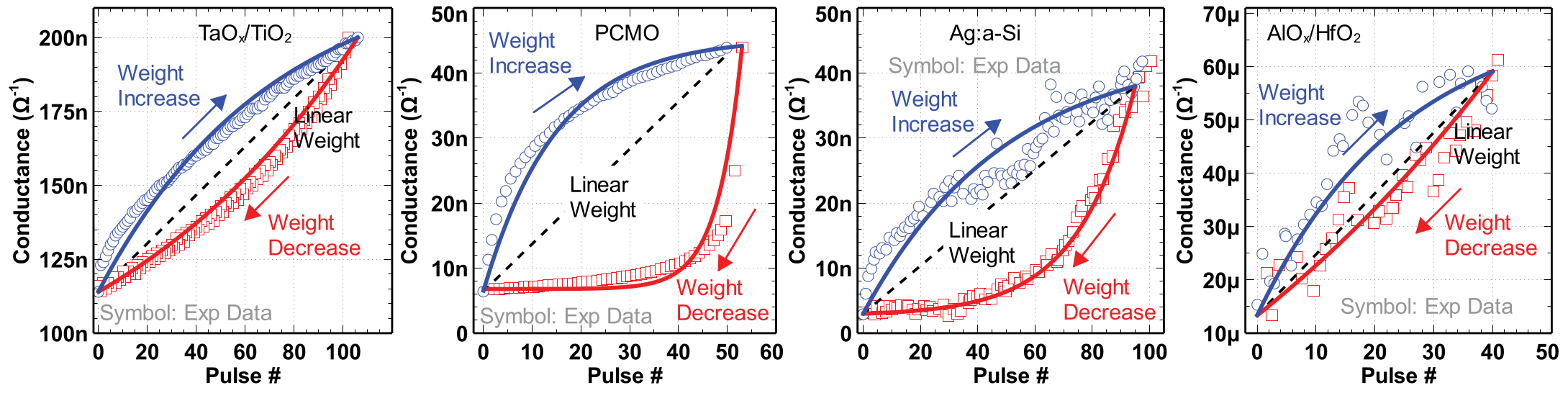}
		\caption{Conductance variation with pulse number corresponding to weight increase and decrease, for $\mathrm{TiO_x/TiO_2}$~\cite{gao2015fully}, PCMO~\cite{park2013neuromorphic}, Ag:a-Si~\cite{jo2010nanoscale}, and $\mathrm{AlO_x/HfO_2}$~\cite{woo2016improved}.}
		\label{Fig:RRAM_Synapse}
	\end{figure}

	Beyond the conventional binary states of \ac{LRS} and \ac{HRS}, \ac{RRAM} demonstrates multiple intermediate controllable resistance states, which shows its potential for emulating synaptic plasticity and implementing analog weight storage in neuromorphic systems. Neural network accuracy is directly related to the number of distinct conductance levels available within the synaptic device. Therefore, artificial synapses capable of achieving numerous intermediate conductance states are highly desirable. Lower precision synaptic weights are preferred to reduce computation and energy consumption, however, the online training of \ac{MNIST} dataset requires at least 6-bit of accuracy~\cite{yu2016binary}. Wu \emph{et.al.} demonstrated $\sim$128 ($2^7$) distinct conductance levels in $\mathrm{HfO_x}$-based \ac{RRAM}, demonstrating that RRAM technology can provide sufficient precision for demanding online learning tasks. A necessary requirement for efficient learning in a neural network is to have linear and symmetric weight update characteristics of the synaptic devices. Linearity allows for precise control of weight changes proportional to applied stimuli, simplifying learning algorithms and enhancing predictability. Symmetric potentiation and depression prevent weight drift, enabling accurate learning of both positive and negative correlations, mirroring biological synapses, and ultimately enhancing network accuracy, and convergence speed. Despite showing hundreds of conducting levels, state-of-the-art RRAM devices show nonlinear and asymmetric weight update (conductance) characteristics while programmed under identical pulses. Figure \ref{Fig:RRAM_Synapse} shows change of conductance with the pulse numbers corresponding to increase of weight (potentiation) and decrease (depression) for $\mathrm{TiO_x/TiO_2}$~\cite{gao2015fully}, PCMO~\cite{park2013neuromorphic}, Ag:a-Si~\cite{jo2010nanoscale}, and $\mathrm{AlO_x/HfO_2}$~\cite{woo2016improved}  
	which shows a highly non-linear and asymmetric nature. Non-linearity in synaptic weight updates negatively impacts network learning and accuracy. It has been shown that the nonlinearity can be improved by either voltage pulses with linearly increasing amplitude~\cite{jang2015optimization}, or nonidentical pulses ~\cite{chen2015mitigating}. A few hardware-efficient algorithms~\cite{chang2017mitigating,lee2022crus} have also been developed to mitigate nonlinearity. \ac{RRAM}'s ability to exhibit a gradual change in conductance allows for the implementation of \ac{STDP} for unsupervised learning in \acp{SNN}. Jo \emph{et.al.}~\cite{jo2010nanoscale} demonstrated the feasibility of using nanoscale memristor devices, including RRAM, as synapses in neuromorphic systems, specifically highlighting their potential for STDP implementation. Specifically, they showed that the resistance of their fabricated memristor device can be modulated based on the relative timing of pre and post-synaptic spikes, mimicking the potentiation and depression noticed in neural connections. From a circuit perspective, a 2T1R artificial synapse has been shown~\cite{wang20152} to achieve STDP by modulating the memristor's conductance based on the temporal relationship between pre-synaptic spikes and post-synaptic spikes. In this work, they also showed the two transistors in the structure not only enable the spike timing comparison but also control the current through the memristor, contributing to stochastic learning due to the device's inherent variability and state-dependent responses.

	\section{Spintronics devices}
	\subsection{Variuos spintronic devices and its operation}
	\begin{figure}[!b]
		\centering
		\includegraphics[scale=0.55]{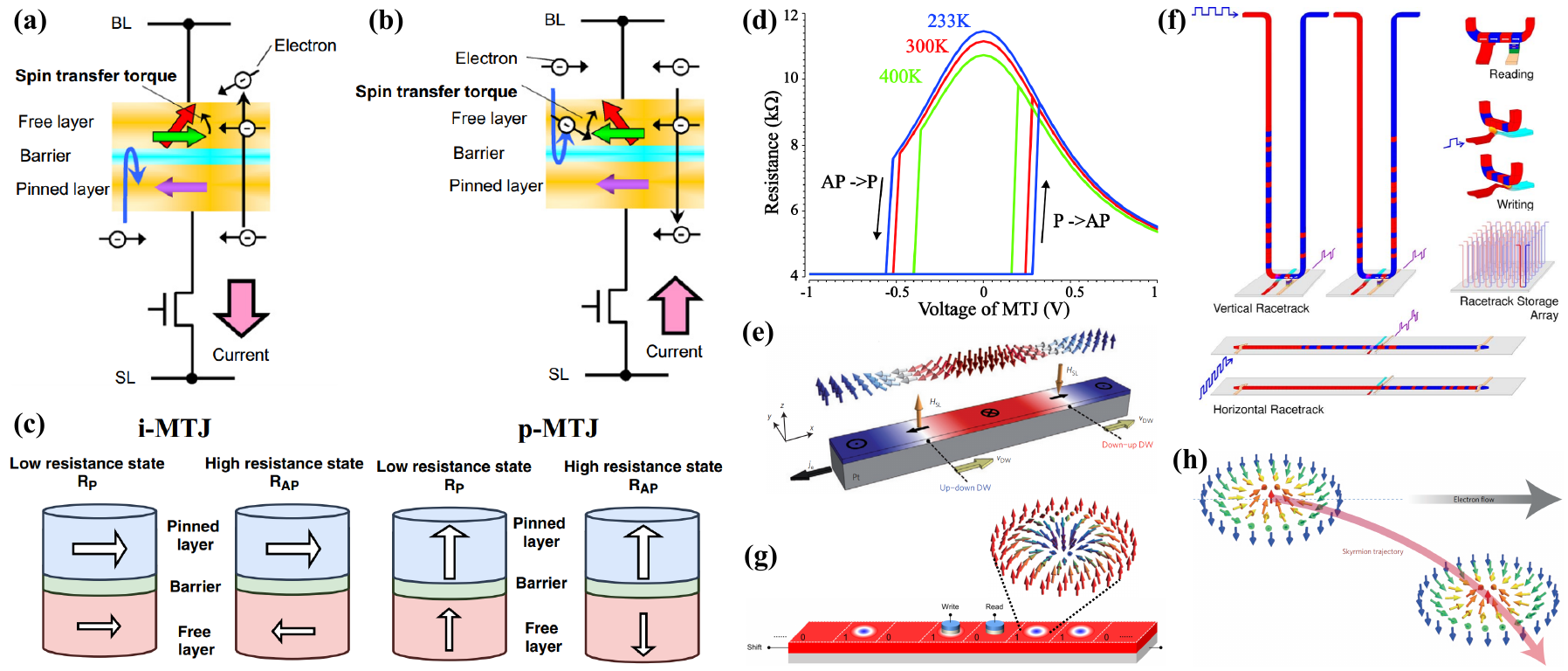}
		\caption{Schematic of \acf{STT} induced switching the MTJ (a) from AP to P state, (b) from P to AP state. Figures are adapted from \cite{kawahara2012spin}. (c) Schematic of in-plane MTJ (i-MTJ) and perpendicular-to-plane MTJ (p-MTJ) with magnetization direction corresponding to pinned and free layer~\cite{barla2021spintronic}. (d) Variation of MTJ resistance with the voltage applied across it, demonstrating the switching from AP to P and P to AP states~\cite{wang2015compact}. (e) Illustration of magnetic \acf{DW} based devices, where arrows point local magnetization direction~\cite{emori2013current}. The white region is the domain wall which acts as an interface between magnetic domains magnetized in opposite directions. (f) Vertical and horizontal racetrack memory based on magnetic \ac{DW}~\cite{parkin2009racetrack}. (f) Magnetic skyrmion-based racetrack device along with read and write head~\cite{luo2021skyrmion,kezsmarki2015neel}. (g) A schematic showing a deflected path traversed by skyrmion due to skyrmion Hall-effect~\cite{chen2017skyrmion}.}
		\label{Fig:MTJ_STT}
	\end{figure}
	In previous sections, we have learned that \ac{PCM} and \ac{RRAM} operate by changing the material's physical structure, which may lead to a lower lifespan for these devices. 
	Spintronic devices offer an alternative paradigm by leveraging the intrinsic spin of electrons, an inherent quantum mechanical property, to store and process information. 
	Giant Magnetoresistance (GMR)~\cite{binasch1989enhanced,baibich1988giant}, discovered in 1988 by Albert Fert and Peter Gr$\mathrm{\ddot{u}}$nberg, marked the inception of the spintronics journey. 
	\ac{GMR} utilizes spin-dependent electron scattering in multilayers, where relative magnetic alignment between the \ac{FL} and the \ac{PL} results in low resistance parallel alignment, whereas anti-parallel alignment shows high resistance. 
	This phenomenon revolutionized hard drive technology by significantly increasing data storage density. 
	\Ac{FL} refers to the ferromagnetic layer whose magnetization can be altered whereas the magnetization of the \ac{PL} is fixed with the help of an antiferromagnet layer, justifying the name. 
	The thickness of the \ac{PL} is made thicker compared to the \ac{FL} so that a larger magnetic/electrical energy is required to change the magnetization direction. 
	In the early days, the magnetization of the \ac{FL} is altered with the help of an external magnetization. With the improvement of the technology, the device becomes smaller to implement denser memory. Due to the unlocalized nature of the magnetic field, this became a bottleneck, showing the requirement of an alternative method to alter the magnetization direction of the \ac{FL}. Also, to enhance the magnetoresistance, researchers replaced the non-magnetic metallic spacer in GMR structures with a thin insulating barrier, enabling spin-dependent electron tunneling, creating \ac{TMR} devices. Such devices are called \ac{MTJ}, where the tunneling probability, and thus resistance, is dictated by the relative magnetization orientations of the FM layers, leading to significantly higher magnetoresistance ratios. \ac{TMR} represents the relative difference between antiparallel ($\mathrm{R_{AP}}$) and parallel ($\mathrm{R_P}$) resistances, normalized with respect to parallel resistance, and is denoted by 
	\begin{equation}
		\mathrm{TMR=\frac{R_{AP}-R_P}{R_P}}.
		\label{Eq:TMR}
	\end{equation}
	As an alternative to switching by a magnetic field, researchers proposed the \ac{STT} mechanism, by utilizing a spin-polarized current to transfer spin angular momentum, enabling the manipulation of the magnetic layers' magnetization dynamics~\cite{ralph2008spin,li2003magnetization,kawahara2012spin}. AP to P switching can be obtained by passing the unpolarized electrons from the \ac{PL} to \ac{FL}, tunneling via the oxide layer as shown in Figure~\ref{Fig:MTJ_STT}(a). Unpolarized electrons entering the \ac{PL} become spin-polarized parallel to its magnetization. These electrons tunnel through the oxide barrier into the \ac{FL}, which is initially in an antiparallel state. Once the spin angular momentum transferred from spin-polarized electrons to the \ac{FL} exceeds a specific critical level, the \ac{FL} magnetization switches to a parallel configuration~\cite{kawahara2012spin}. On the other hand, the switching from P state to AP state is done by sending the current in the opposite direction, \emph{i.e.} now the electrons enter from \ac{FL}, as shown in Figure \ref{Fig:MTJ_STT}(b). When electrons traverse the \ac{FL}, those with spin aligned to the \ac{PL}'s magnetization pass through. Electrons with opposite spin are reflected at the oxide barrier and injected back into the \ac{FL}. This reflected, spin-polarized current exerts \ac{STT} on the \ac{FL}'s magnetization, causing it to switch when the current surpasses a critical threshold. The \ac{FL}'s magnetic configuration is governed by uniaxial anisotropy, establishing two energy minima corresponding to magnetization states oriented $180^{\circ}$ apart. The magnetization of the \ac{FL} can be either oriented within the film plane, known as in-plane MTJ (i-MTJ), or perpendicular to the film plane, known as p-MTJ as shown in Figure \ref{Fig:MTJ_STT}(c). Between these two types of \acp{MTJ}, p-MTJ is preferred due to the lower critical current needed than the i-MTJ~\cite{yoda2012progress}. \Ac{SOT} offers a promising alternative to \ac{STT} by enabling magnetization switching with potentially lower energy consumption due to its separate charge and magnetic current pathways. The memory stored in MTJ devices is binary, represented by the P or AP state of magnetic layers. However, to move beyond simple binary storage, researchers have explored magnetic domain walls~\cite{emori2013current}, where, instead of uniform alignment, the magnetization undergoes a gradual shift between two opposing orientations dictated by uniaxial anisotropy, as shown in Figure \ref{Fig:MTJ_STT}(e). Here, arrows pointing to the local magnetization direction (up and down in the figure) and the gradual transition shown by the white region, known as the \acf{DW}, have the local magnetization along the plane. Parkin \emph{et.al.} demonstrated such devices can encode multiple information bits stored and accessed in a racetrack-like structure~\cite{parkin2008magnetic,parkin2009racetrack} as shown in Figure \ref{Fig:MTJ_STT}(f). Driven by the need for smaller size and lower energy consumption, researchers have developed magnetic skyrmion-based devices~\cite{woo2016observation,yu2016room}, which exhibit significantly lower depinning current densities~\cite{jonietz2010spin} compared to conventional domain walls, making them promising candidates for ultra-low power spintronic applications. Magnetic skyrmions are nanoscale, vortex-like magnetic textures that exhibit unique topological properties. Unlike conventional magnetic domains, skyrmions possess a swirling spin structure, where the magnetization rotates in a chiral manner around a central core. This vortex-like configuration forms a topologically protected spin texture, meaning it is robust against continuous deformations and defects. The core of the skyrmion exhibits a magnetization direction opposite to its surroundings, forming a characteristic circular spin pattern, as depicted in Figure \ref{Fig:MTJ_STT}(g). Thanks to their unique magnetic texture, skyrmions can be manipulated with spin currents or magnetic fields, making them suitable for spintronic memory and logic in dense, energy-saving systems.
	However, despite their promising attributes, skyrmions suffer from the skyrmion Hall effect~\cite{chen2017skyrmion}, causing them to deviate from a straight path, as shown in Figure \ref{Fig:MTJ_STT}(h), which complicates precise control and device implementation. Zhang \emph{et.al.} demonstrated instead of taking a single thin film of nanomagnetic layer, two magnetic thin fils with opposite magnetization coupled via an antiferromagnetic coupling, can mitigate such effect, thus enhancing the speed of the device drastically~\cite{zhang2016magnetic}. Skyrmion-based devices encode binary information through the presence or absence of skyrmions, and multiple skyrmions within a racetrack structure can represent a binary data stream.
	
	The magnetization dynamics of any magnetic system be it \ac{MTJ}, DW or skyrmion, is calculated using \ac{LLGS} equation~\cite{slonczewski1989conductance,sengupta2016short}
	\begin{equation}
		\left(1+\alpha^2\right)\frac{d\hat{\mathbf{m}}}{dt}=-\gamma\left(\hat{\mathbf{m}}\times \mathbf{H}_{eff}\right) - \alpha\gamma\left(\hat{\mathbf{m}}\times\hat{\mathbf{m}}\times \mathbf{H}_{eff}\right) + \frac{1}{qN_S}\left(\hat{\mathbf{m}}\times\mathbf{I}_S\times\hat{\mathbf{m}}\right)
		\label{LLGS}.
	\end{equation}
	Here, $\hat{\mathbf{m}}$ denotes the unit vector representing the magnetization of \ac{FL}, $\gamma$ denotes the gyromagnetic ratio, $\alpha$ is the damping factor, $\mathbf{H}_{eff}$ is the effective magnetic field, that consists of uniaxial anisotropy field, demagnetization field. Thermal noise is included as an additional thermal field in the effective field, given by 
	\begin{equation}
		\mathbf{H}_{th}=\sqrt{\frac{\alpha}{1+\alpha^2}\frac{2K_B T}{\gamma \mu_0 M_S V \delta_t}}G_{0,1}. 
	\end{equation}
	where $K_B$ is the Boltzmann constant, $T$ is the temperature, $\mu$ free space permeability, $M_S$ is the saturation magnetization, $V$ volume of the \ac{FL}, $\delta_t$ is the time step for the simulation, and $G_{0,1}$ is Gaussian distribution with zero mean, and unit standard deviation. Other terms in Eq.~(\ref{LLGS}), such as $q$ is the charge of electron, $N_S=(M_S V/\mu_B)$ is the number of spins in the \ac{FL}, $\mu_B$ is Bohr magneton, and $\mathbf{I}_S$ spin current. 
	
	\subsection{Spitronic devices as Neuron}
	\begin{figure}[!b]
		\centering
		\includegraphics[scale=0.75]{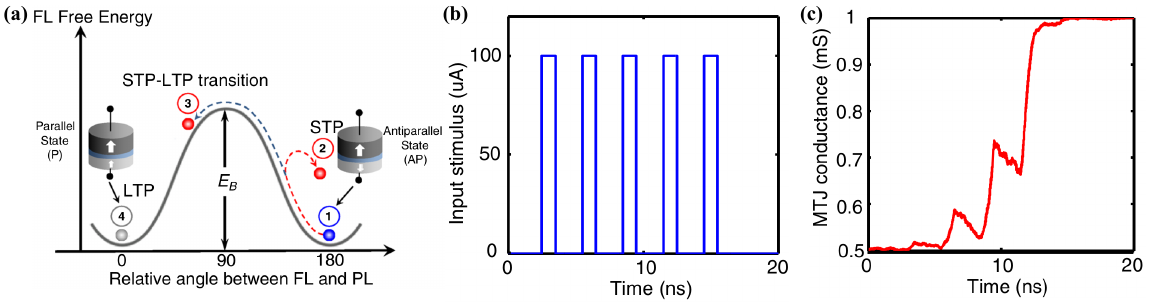}
		\caption{(a) FL free energy profile against the relative angle between the FL and PL, demonstrating the energy barrier $E_B$ two stable state P and AP. (b) Short current pulses that are applied across the MTJ, and (c) corresponding change of the conductance of the MTJ. Figures are adapted from Ref~\cite{sengupta2016short}.}
		\label{Fig:MTJ_pulse}
	\end{figure}
	In \acp{SNN}, spintronic devices, ranging from MTJs to skyrmion-based structures, have been used to implement both \ac{IF} and \ac{LIF} neuron models. The stability of magnetization along a nanomagnet's easy axis, determined by its uniaxial anisotropy energy barrier, necessitates the application of a current pulse with sufficient amplitude and duration to induce state transitions, such as from anti-parallel (AP) to parallel (P) and vice versa. Figure \ref{Fig:MTJ_pulse}(a) the \ac{FL} free energy profile with the relative angle between the magnetization of the \ac{PL} and \ac{FL}. It shows two valleys, \emph{i.e.} the stable states, are at $180^{\circ}$ apart with an energy barrier of height $E_B=K_B T$. Thus, switching from AP state (position \circled{1}) to P state (position \circled{4}) or vice versa we need to apply sufficient current to cross the energy barrier. This energy barrier characteristic can be leveraged to introduce leak functionality by applying current pulses with durations \emph{insufficient} to fully overcome the barrier, thereby inducing partial state transitions. Figure \ref{Fig:MTJ_pulse}(b) shows such current pulses applied to the \ac{MTJ}. Each pulse attempts to rotate the free layer's magnetization, pushing the state from its stable valley towards the energy barrier peak. If the pulse width is insufficient, the state will relax back towards its stable state before the arrival of the next pulse, as shown in the transition from position \circled{2} to position \circled{1}. The conductance of the MTJ is dependent on the relative angle between the magnetization of the PL and the FL layer as given by~\cite{sengupta2016short}
	\begin{equation}
		G = G_{P} cos^2\left(\frac{\theta}{2}\right) + G_{AP} sin^2\left(\frac{\theta}{2}\right)
		\label{Eq_MTJ_G}
	\end{equation} 
	where, $G_P\left(G_{AP}\right)$ is the conductance of the MTJ in P(AP) state.
	If initially the \ac{MTJ} is in AP state, \emph{i.e.} the angle between the magnetization of the \ac{FL} and \ac{PL} is $180^{\circ}$, leading to $G=G_{AP}$, a low value of conductance. In the presence of such short pulses, the magnetization vector ($\hat{\mathbf{m}}$) moves away from the AP state, leading to the reduction of the relative angle $\theta$, and according to Eq. (\ref{Eq_MTJ_G}), conductance increases. This is equivalent to the integration functionality of the biological neuron. After the pulse is removed $\hat{\mathbf{m}}$ starts to move towards the easy axis due to the uniaxial anisotropy field, which increases $\theta$, leading to a reduction of the conductance, thus imitating leak functionality. If a sufficient number of pulses are applied across the MTJ and switched to the P state ($\theta=0^{\circ}$), it achieves the highest conductance, \emph{i.e.} $G=G_P$. This can be utilized to incorporate firing by adding an extra circuitry in series with the MTJ. \ac{SOT}-based MTJ is also used to immitate \ac{LIF} neuron~\cite{sengupta2016magnetic}, as it decouples the read and write path. Apart from MTJ, spintronic devices with magnetic textures such as \ac{DW} and skyrmion-based devices can also be used to emulate the neuronal functionality. The domain wall or skyrmion's position could be used to represent the neuron's membrane potential. With the application of current pulses DW and skyrmion move along the nanotrack, imitating the integration functionality, and upon reaching the end of the nanotrack, a firing spike can be generated with the help of external circuitry, by sensing the DW or skyrmion using a MTJ like sensor~\cite{das2022self,das2023bilayer}. The leak functionality in such devices is incorporated either by introducing uniaxial anisotropy gradient~\cite{das2023bilayer} or, by modifying the shape of the nanotrack~\cite{brigner2019shape}. 
	
	\subsection{Spintronic Devices as Synapse}
\begin{figure}[!h]
		\centering
		\includegraphics[scale=0.6]{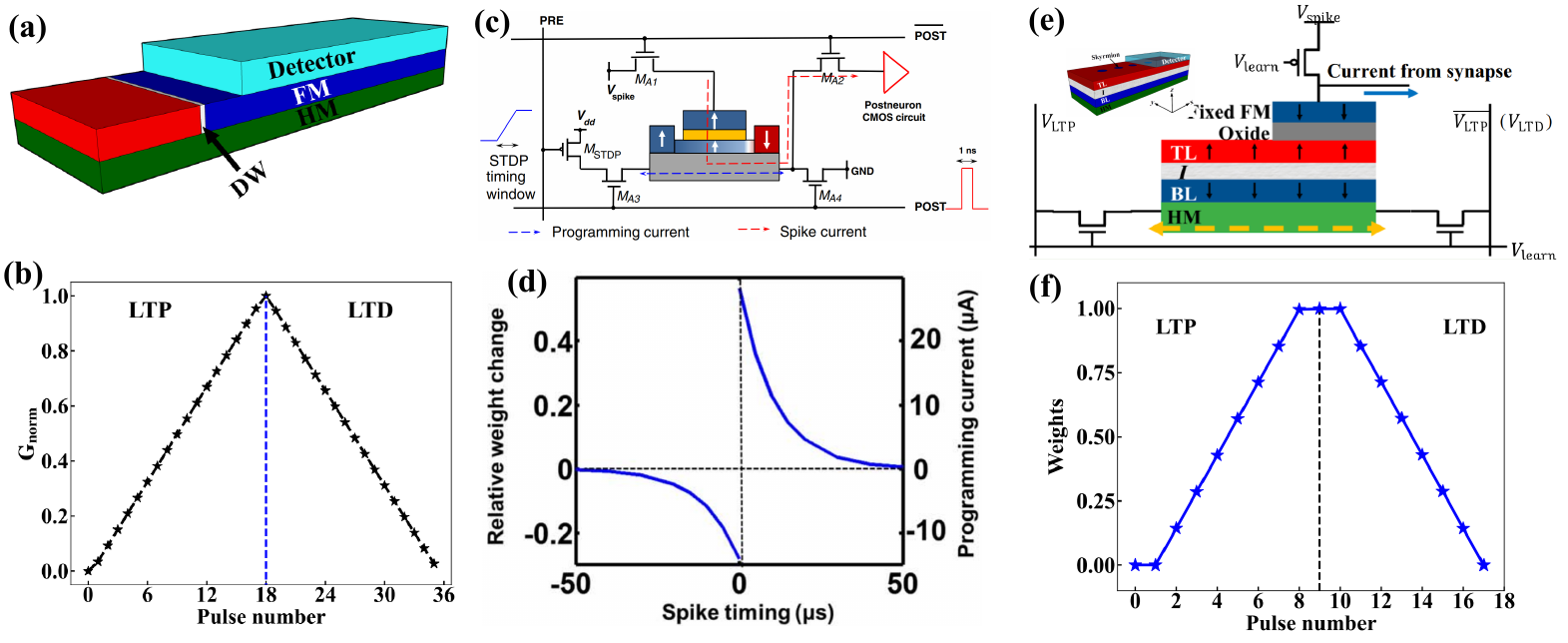}
		\caption{(a) Schematic of the DW based synapse. (b) Variation of conductance with the pulse number for the \ac{LTP} and \ac{LTD}~\cite{das2022design}. (c) Circuit showing the implementation of the \ac{STDP} and (d) relative weight change with spike timing~\cite{sengupta2016hybrid}. (e) Implementation of the skyrmion-based synaptic device, and (f) Variation of weight (conductance) with the pulse number for the \ac{LTP} and \ac{LTD}~\cite{das2023bilayer}. }
		\label{Fig:Spin_Synapse}
	\end{figure}

	Spintronic devices, beyond their applications in memory and logic, offer a compelling pathway for emulating synaptic functionality in neuromorphic systems. Their ability to precisely control magnetic states through spin currents, combined with inherent non-volatility, makes them attractive candidates for building energy-efficient and adaptable artificial synapses. By leveraging phenomena like \ac{STT}, and \ac{SOT}, researchers are exploring novel spintronic synapse designs capable of mimicking the plasticity and learning capabilities of biological synapses. Synapses demand multilevel storage to accurately represent varying connection strengths. However, traditional \acp{MTJ} inherently stores information in a binary fashion, limiting their direct application in emulating analog synaptic behavior. While digital synaptic implementations can be achieved by combining multiple MTJs to represent quantized weight values, this approach deviates from the analog nature of biological synapses. In contrast, magnetic \ac{DW} based devices offer a more biologically plausible solution. By encoding synaptic weight as the position of a domain wall, they enable a continuous range of resistance states, effectively mimicking the analog memory observed in the brain. This gradual transition of magnetization within the domain wall allows for a more nuanced representation of synaptic weights, facilitating the implementation of analog learning rules and potentially leading to more energy-efficient and biologically accurate neuromorphic systems. For such devices, a longer nanotrack is used, where the DW is driven by external currents similar to synaptic pulses. On top of this, an MTJ-like detector is fabricated where the nanotrack is used as the free layer. The resistance of the detector is measured by passing a small current through the detector (Figure \ref{Fig:Spin_Synapse}(a)). When the magnetization of the PL and FL are antiparallel it gives minimum conductance. Applying positive (negative) current pulses moves the DW in a forward (backward) direction, thus changing the effective resistance measured through a detector. 
	Such variations of the conductance for the \ac{LTP} and \ac{LTD} are shown in Figure \ref{Fig:Spin_Synapse}(b). \ac{SNN} uses \ac{STDP} mechanism for unsupervised learning. A circuit implementation of such learning is demonstrated by Sengupta \emph{et.al.}~\cite{sengupta2016hybrid} for the DW-based devices, as shown in Figure \ref{Fig:Spin_Synapse}(c), and the corresponding temporal weight change is shown in Figure \ref{Fig:Spin_Synapse}(d). Das \emph{et.al.}~\cite{das2023bilayer} demonstrated a similar feature using bilayer skyrmion-based devices. Here instead of taking a single skyrmion, they used 7 skyrmions to implement a total of 3-bit synapses, where sending (removing) synapses under the detector increases (decreases) the conductance of the device. They demonstrated the energy consumption to change a single bit in on the order of a few fJ, indicating a promising potential for ultra-low power neuromorphic systems. Compared to the previously discussed eNVM devices, a notable advantage of DW and skyrmion-based synaptic devices lies in their highly linear and symmetric weight update characteristics under identical current pulses, as evidenced in Figure~\ref{Fig:Spin_Synapse}(b) and (f). This linear and symmetric weight update characteristic is essential for the precise training of neural networks.   
	
	\section{Conclusion}
	In the pursuit of hardware-based neuromorphic systems, PCM, RRAM, and spintronic devices each offer unique strengths and challenges. PCM provides multilevel storage and relatively mature fabrication but suffers from drift and high programming currents. RRAM excels in low-power operation and high switching speeds, offering versatile synaptic plasticity, yet faces challenges in variability and endurance. Spintronics, with its inherent non-volatility and potential for ultra-low power consumption, particularly in skyrmion-based devices, presents a promising avenue for future neuromorphic architectures, although it is still in its early stages of development. While PCM and RRAM offer near-term solutions, spintronics holds the potential for revolutionary advancements in energy-efficient and scalable neuromorphic computing, leading to the next generation of artificial intelligence. The optimal choice depends on application specific requirements, and balancing factors such as energy efficiency, speed, endurance, and fabrication complexity.
	
	\bibliographystyle{ieeetr} 
	\bibliography{References}

@article{maass1997networks,
	title={Networks of spiking neurons: the third generation of neural network models},
	author={Maass, Wolfgang},
	journal={Neural networks},
	volume={10},
	number={9},
	pages={1659--1671},
	year={1997},
	publisher={Elsevier}
}

@article{popescu2009multilayer,
	title={Multilayer perceptron and neural networks},
	author={Popescu, Marius-Constantin and Balas, Valentina E and Perescu-Popescu, Liliana and Mastorakis, Nikos},
	journal={WSEAS Transactions on Circuits and Systems},
	volume={8},
	number={7},
	pages={579--588},
	year={2009},
	publisher={World Scientific and Engineering Academy and Society (WSEAS) Stevens Point~…}
}

@article{kim2002fully,
	title={Fully complex multi-layer perceptron network for nonlinear signal processing},
	author={Kim, Taehwan and Adali, T{\"u}lay},
	journal={Journal of VLSI signal processing systems for signal, image and video technology},
	volume={32},
	pages={29--43},
	year={2002},
	publisher={Springer}
}

@inproceedings{teow2000handwritten,
	title={Handwritten digit recognition with a novel vision model that extracts linearly separable features},
	author={Teow, Loo-Nin and Loe, Kia-Fock},
	booktitle={Proceedings IEEE Conference on Computer Vision and Pattern Recognition. CVPR 2000 (Cat. No. PR00662)},
	volume={2},
	pages={76--81},
	year={2000},
	organization={IEEE}
}

@article{liu2003handwritten,
	title={Handwritten digit recognition: benchmarking of state-of-the-art techniques},
	author={Liu, Cheng-Lin and Nakashima, Kazuki and Sako, Hiroshi and Fujisawa, Hiromichi},
	journal={Pattern recognition},
	volume={36},
	number={10},
	pages={2271--2285},
	year={2003},
	publisher={Elsevier}
}

@article{phung2007pyramidal,
	title={A pyramidal neural network for visual pattern recognition},
	author={Phung, Son Lam and Bouzerdoum, Abdesselam},
	journal={IEEE transactions on neural networks},
	volume={18},
	number={2},
	pages={329--343},
	year={2007},
	publisher={IEEE}
}

@article{chen1996speech,
	title={A speech recognition method based on the sequential multi-layer perceptrons},
	author={Chen, Wen-Yuan and Chen, Sin-Horng and Lin, Cheng-Jung},
	journal={Neural Networks},
	volume={9},
	number={4},
	pages={655--669},
	year={1996},
	publisher={Elsevier}
}

@inproceedings{ahad2002speech,
	title={Speech recognition using multilayer perceptron},
	author={Ahad, Abdul and Fayyaz, Ahsan and Mehmood, Tariq},
	booktitle={IEEE Students Conference, ISCON'02. Proceedings.},
	volume={1},
	pages={103--109},
	year={2002},
	organization={IEEE}
}

@inproceedings{ma2002natural,
	title={Natural language processing with neural networks},
	author={Ma, Qing},
	booktitle={Language engineering conference, 2002. proceedings},
	pages={45--56},
	year={2002},
	organization={IEEE}
}

@article{goldberg2016primer,
	title={A primer on neural network models for natural language processing},
	author={Goldberg, Yoav},
	journal={Journal of Artificial Intelligence Research},
	volume={57},
	pages={345--420},
	year={2016}
}

@inproceedings{shiblee2009time,
	title={Time series prediction with multilayer perceptron (MLP): A new generalized error based approach},
	author={Shiblee, Md and Kalra, Prem Kumar and Chandra, B},
	booktitle={Advances in Neuro-Information Processing: 15th International Conference, ICONIP 2008, Auckland, New Zealand, November 25-28, 2008, Revised Selected Papers, Part II 15},
	pages={37--44},
	year={2009},
	organization={Springer}
}

@article{li2017convergence,
	title={Convergence analysis of two-layer neural networks with relu activation},
	author={Li, Yuanzhi and Yuan, Yang},
	journal={Advances in neural information processing systems},
	volume={30},
	year={2017}
}

@article{lecun1998gradient,
	title={Gradient-based learning applied to document recognition},
	author={LeCun, Yann and Bottou, L{\'e}on and Bengio, Yoshua and Haffner, Patrick},
	journal={Proceedings of the IEEE},
	volume={86},
	number={11},
	pages={2278--2324},
	year={1998},
	publisher={Ieee}
}

@article{krizhevsky2012imagenet,
	title={Imagenet classification with deep convolutional neural networks},
	author={Krizhevsky, Alex and Sutskever, Ilya and Hinton, Geoffrey E},
	journal={Advances in neural information processing systems},
	volume={25},
	year={2012}
}

@article{zhao2017object,
	title={Object-based convolutional neural network for high-resolution imagery classification},
	author={Zhao, Wenzhi and Du, Shihong and Emery, William J},
	journal={IEEE Journal of Selected Topics in Applied Earth Observations and Remote Sensing},
	volume={10},
	number={7},
	pages={3386--3396},
	year={2017},
	publisher={IEEE}
}

@inproceedings{ren2016single,
	title={Single image dehazing via multi-scale convolutional neural networks},
	author={Ren, Wenqi and Liu, Si and Zhang, Hua and Pan, Jinshan and Cao, Xiaochun and Yang, Ming-Hsuan},
	booktitle={Computer Vision--ECCV 2016: 14th European Conference, Amsterdam, The Netherlands, October 11-14, 2016, Proceedings, Part II 14},
	pages={154--169},
	year={2016},
	organization={Springer}
}

@article{davies2018loihi,
	title={Loihi: A neuromorphic manycore processor with on-chip learning},
	author={Davies, Mike and Srinivasa, Narayan and Lin, Tsung-Han and Chinya, Gautham and Cao, Yongqiang and Choday, Sri Harsha and Dimou, Georgios and Joshi, Prasad and Imam, Nabil and Jain, Shweta and others},
	journal={Ieee Micro},
	volume={38},
	number={1},
	pages={82--99},
	year={2018},
	publisher={IEEE}
}

@inproceedings{cassidy202411,
	title={11.4 IBM NorthPole: An Architecture for Neural Network Inference with a 12nm Chip},
	author={Cassidy, Andrew S and Arthur, John V and Akopyan, Filipp and Andreopoulos, Alexander and Appuswamy, Rathinakumar and Datta, Pallab and Debole, Michael V and Esser, Steven K and Otero, Carlos Ortega and Sawada, Jun and others},
	booktitle={2024 IEEE International Solid-State Circuits Conference (ISSCC)},
	volume={67},
	pages={214--215},
	year={2024},
	organization={IEEE}
}

@inproceedings{kamata2022fully,
	title={Fully spiking variational autoencoder},
	author={Kamata, Hiromichi and Mukuta, Yusuke and Harada, Tatsuya},
	booktitle={Proceedings of the AAAI conference on artificial intelligence},
	volume={36},
	number={6},
	pages={7059--7067},
	year={2022}
}

@article{zhou2021improved,
	title={Improved integrate-and-fire neuron models for inference acceleration of spiking neural networks},
	author={Zhou, Yongcheng and Zhang, Anguo},
	journal={Applied Intelligence},
	volume={51},
	number={4},
	pages={2393--2405},
	year={2021},
	publisher={Springer}
}

@book{nielsen2015,
	title     = {Neural Networks and Deep Learning},
	author    = {Nielsen, Michael},
	year      = 2015,
	note      = {\url{http://neuralnetworksanddeeplearning.com/}}
}

@article{lillicrap2020backpropagation,
	title={Backpropagation and the brain},
	author={Lillicrap, Timothy P and Santoro, Adam and Marris, Luke and Akerman, Colin J and Hinton, Geoffrey},
	journal={Nature Reviews Neuroscience},
	volume={21},
	number={6},
	pages={335--346},
	year={2020},
	publisher={Nature Publishing Group UK London}
}

@article{bi2001synaptic,
	title={Synaptic modification by correlated activity: Hebb's postulate revisited},
	author={Bi, Guo-qiang and Poo, Mu-ming},
	journal={Annual review of neuroscience},
	volume={24},
	number={1},
	pages={139--166},
	year={2001},
	publisher={Annual Reviews 4139 El Camino Way, PO Box 10139, Palo Alto, CA 94303-0139, USA}
}

@article{cai2014neuronal,
	title={Neuronal synapse as a memristor: Modeling pair-and triplet-based STDP rule},
	author={Cai, Weiran and Ellinger, Frank and Tetzlaff, Ronald},
	journal={IEEE transactions on biomedical circuits and systems},
	volume={9},
	number={1},
	pages={87--95},
	year={2014},
	publisher={IEEE}
}

@article{stern2021uncovering,
	title={Uncovering Phase Change Memory Energy Limits by Sub-Nanosecond Probing of Power Dissipation Dynamics},
	author={Stern, Keren and Wainstein, Nicol{\'a}s and Keller, Yair and Neumann, Christopher M and Pop, Eric and Kvatinsky, Shahar and Yalon, Eilam},
	journal={Advanced Electronic Materials},
	volume={7},
	number={8},
	pages={2100217},
	year={2021},
	publisher={Wiley Online Library}
}

@inproceedings{jeyasingh2012phase,
	title={Phase change memory: Scaling and applications},
	author={Jeyasingh, Rakesh and Liang, Jiale and Caldwell, Marissa A and Kuzum, Duygu and Wong, H-S Philip},
	booktitle={Proceedings of the IEEE 2012 Custom Integrated Circuits Conference},
	pages={1--7},
	year={2012},
	organization={IEEE}
}

@article{tuma2016stochastic,
	title={Stochastic phase-change neurons},
	author={Tuma, Tomas and Pantazi, Angeliki and Le Gallo, Manuel and Sebastian, Abu and Eleftheriou, Evangelos},
	journal={Nature nanotechnology},
	volume={11},
	number={8},
	pages={693--699},
	year={2016},
	publisher={Nature Publishing Group UK London}
}

@article{cobley2018self,
	title={A self-resetting spiking phase-change neuron},
	author={Cobley, RA and Hayat, H and Wright, CD},
	journal={Nanotechnology},
	volume={29},
	number={19},
	pages={195202},
	year={2018},
	publisher={IOP Publishing}
}

@article{kuzum2012nanoelectronic,
	title={Nanoelectronic programmable synapses based on phase change materials for brain-inspired computing},
	author={Kuzum, Duygu and Jeyasingh, Rakesh GD and Lee, Byoungil and Wong, H-S Philip},
	journal={Nano letters},
	volume={12},
	number={5},
	pages={2179--2186},
	year={2012},
	publisher={ACS Publications}
}

@article{bi1998synaptic,
	title={Synaptic modifications in cultured hippocampal neurons: dependence on spike timing, synaptic strength, and postsynaptic cell type},
	author={Bi, Guo-qiang and Poo, Mu-ming},
	journal={Journal of neuroscience},
	volume={18},
	number={24},
	pages={10464--10472},
	year={1998},
	publisher={Society for Neuroscience}
}

@article{ambrogio2016unsupervised,
	title={Unsupervised learning by spike timing dependent plasticity in phase change memory (PCM) synapses},
	author={Ambrogio, Stefano and Ciocchini, Nicola and Laudato, Mario and Milo, Valerio and Pirovano, Agostino and Fantini, Paolo and Ielmini, Daniele},
	journal={Frontiers in neuroscience},
	volume={10},
	pages={56},
	year={2016},
	publisher={Frontiers Media SA}
}

@article{ielmini2012evidence,
	title={Evidence for voltage-driven set/reset processes in bipolar switching RRAM},
	author={Ielmini, Daniele and Nardi, Federico and Balatti, Simone},
	journal={IEEE Transactions on Electron Devices},
	volume={59},
	number={8},
	pages={2049--2056},
	year={2012},
	publisher={IEEE}
}

@article{yu2013hfox,
	title={HfOx-based vertical resistive switching random access memory suitable for bit-cost-effective three-dimensional cross-point architecture},
	author={Yu, Shimeng and Chen, Hong-Yu and Gao, Bin and Kang, Jinfeng and Wong, H-S Philip},
	journal={ACS nano},
	volume={7},
	number={3},
	pages={2320--2325},
	year={2013},
	publisher={ACS Publications}
}

@article{chu2013charge,
	title={Charge quantity influence on resistance switching characteristic during forming process},
	author={Chu, Tian-Jian and Chang, Ting-Chang and Tsai, Tsung-Ming and Wu, Hsing-Hua and Chen, Jung-Hui and Chang, Kuan-Chang and Young, Tai-Fa and Chen, Kai-Hsang and Syu, Yong-En and Chang, Geng-Wei and others},
	journal={IEEE Electron Device Letters},
	volume={34},
	number={4},
	pages={502--504},
	year={2013},
	publisher={IEEE}
}

@article{liu2019resistive,
	title={Resistive switching characteristics of HfO2 thin films on mica substrates prepared by sol-gel process},
	author={Liu, Chao-Feng and Tang, Xin-Gui and Wang, Lun-Quan and Tang, Hui and Jiang, Yan-Ping and Liu, Qiu-Xiang and Li, Wen-Hua and Tang, Zhen-Hua},
	journal={Nanomaterials},
	volume={9},
	number={8},
	pages={1124},
	year={2019},
	publisher={MDPI}
}

@article{lashkare2018pcmo,
	title={PCMO RRAM for integrate-and-fire neuron in spiking neural networks},
	author={Lashkare, Sandip and Chouhan, S and Chavan, T and Bhat, A and Kumbhare, P and Ganguly, U},
	journal={IEEE Electron Device Letters},
	volume={39},
	number={4},
	pages={484--487},
	year={2018},
	publisher={IEEE}
}

@article{bian2022stacked,
	title={A stacked memristive device enabling both analog and threshold switching behaviors for artificial leaky integrate and fire neuron},
	author={Bian, Jingyao and Tao, Ye and Wang, Zhongqiang and Zhang, Xiaohan and Zhao, Xiaoning and Lin, Ya and Xu, Haiyang and Liu, Yichun},
	journal={IEEE Electron Device Letters},
	volume={43},
	number={9},
	pages={1436--1439},
	year={2022},
	publisher={IEEE}
}

@inproceedings{yu2016binary,
	title={Binary neural network with 16 Mb RRAM macro chip for classification and online training},
	author={Yu, Shimeng and Li, Zhiwei and Chen, Pai-Yu and Wu, Huaqiang and Gao, Bin and Wang, Deli and Wu, Wei and Qian, He},
	booktitle={2016 IEEE International Electron Devices Meeting (IEDM)},
	pages={16--2},
	year={2016},
	organization={IEEE}
}

@article{gao2015fully,
	title={Fully parallel write/read in resistive synaptic array for accelerating on-chip learning},
	author={Gao, Ligang and Wang, I-Ting and Chen, Pai-Yu and Vrudhula, Sarma and Seo, Jae-sun and Cao, Yu and Hou, Tuo-Hung and Yu, Shimeng},
	journal={Nanotechnology},
	volume={26},
	number={45},
	pages={455204},
	year={2015},
	publisher={IOP Publishing}
}

@inproceedings{park2013neuromorphic,
	title={Neuromorphic speech systems using advanced ReRAM-based synapse},
	author={Park, S and Sheri, A and Kim, J and Noh, J and Jang, J and Jeon, M and Lee, B and Lee, BR and Lee, BH and Hwang, H-J},
	booktitle={2013 IEEE International Electron Devices Meeting},
	pages={25--6},
	year={2013},
	organization={IEEE}
}

@article{jo2010nanoscale,
	title={Nanoscale memristor device as synapse in neuromorphic systems},
	author={Jo, Sung Hyun and Chang, Ting and Ebong, Idongesit and Bhadviya, Bhavitavya B and Mazumder, Pinaki and Lu, Wei},
	journal={Nano letters},
	volume={10},
	number={4},
	pages={1297--1301},
	year={2010},
	publisher={ACS Publications}
}

@article{woo2016improved,
	title={Improved synaptic behavior under identical pulses using AlO x/HfO 2 bilayer RRAM array for neuromorphic systems},
	author={Woo, Jiyong and Moon, Kibong and Song, Jeonghwan and Lee, Sangheon and Kwak, Myounghun and Park, Jaesung and Hwang, Hyunsang},
	journal={IEEE Electron Device Letters},
	volume={37},
	number={8},
	pages={994--997},
	year={2016},
	publisher={IEEE}
}

@article{lee2022crus,
	title={CRUS: A hardware-efficient algorithm mitigating highly nonlinear weight update in CIM crossbar arrays for artificial neural networks},
	author={Lee, Junmo and Hwang, Joon and Cho, Youngwoon and Park, Min-Kyu and Choi, Woo Young and Kim, Sangbum and Lee, Jong-Ho},
	journal={IEEE Journal on Exploratory Solid-State Computational Devices and Circuits},
	volume={8},
	number={2},
	pages={145--154},
	year={2022},
	publisher={IEEE}
}

@article{chang2017mitigating,
	title={Mitigating asymmetric nonlinear weight update effects in hardware neural network based on analog resistive synapse},
	author={Chang, Chih-Cheng and Chen, Pin-Chun and Chou, Teyuh and Wang, I-Ting and Hudec, Boris and Chang, Che-Chia and Tsai, Chia-Ming and Chang, Tian-Sheuan and Hou, Tuo-Hung},
	journal={IEEE Journal on Emerging and Selected Topics in Circuits and Systems},
	volume={8},
	number={1},
	pages={116--124},
	year={2017},
	publisher={IEEE}
}

@article{jang2015optimization,
	title={Optimization of conductance change in Pr 1--x Ca x MnO 3-based synaptic devices for neuromorphic systems},
	author={Jang, Jun-Woo and Park, Sangsu and Burr, Geoffrey W and Hwang, Hyunsang and Jeong, Yoon-Ha},
	journal={IEEE Electron Device Letters},
	volume={36},
	number={5},
	pages={457--459},
	year={2015},
	publisher={IEEE}
}

@inproceedings{chen2015mitigating,
	title={Mitigating effects of non-ideal synaptic device characteristics for on-chip learning},
	author={Chen, Pai-Yu and Lin, Binbin and Wang, I-Ting and Hou, Tuo-Hung and Ye, Jieping and Vrudhula, Sarma and Seo, Jae-sun and Cao, Yu and Yu, Shimeng},
	booktitle={2015 IEEE/ACM International Conference on Computer-Aided Design (ICCAD)},
	pages={194--199},
	year={2015},
	organization={IEEE}
}

@article{wang20152,
	title={A 2-transistor/1-resistor artificial synapse capable of communication and stochastic learning in neuromorphic systems},
	author={Wang, Zhongqiang and Ambrogio, Stefano and Balatti, Simone and Ielmini, Daniele},
	journal={Frontiers in neuroscience},
	volume={8},
	pages={438},
	year={2015},
	publisher={Frontiers Media SA}
}

@article{binasch1989enhanced,
	title={Enhanced magnetoresistance in layered magnetic structures with antiferromagnetic interlayer exchange},
	author={Binasch, Gr{\"u}nberg and Gr{\"u}nberg, Peter and Saurenbach, F and Zinn, W},
	journal={Physical review B},
	volume={39},
	number={7},
	pages={4828},
	year={1989},
	publisher={APS}
}

@article{baibich1988giant,
	title={Giant magnetoresistance of (001) Fe/(001) Cr magnetic superlattices},
	author={Baibich, Mario Norberto and Broto, Jean Marc and Fert, Albert and Van Dau, F Nguyen and Petroff, Fr{\'e}d{\'e}ric and Etienne, P and Creuzet, G and Friederich, A and Chazelas, J},
	journal={Physical review letters},
	volume={61},
	number={21},
	pages={2472},
	year={1988},
	publisher={APS}
}

@article{ralph2008spin,
	title={Spin transfer torques},
	author={Ralph, Daniel C and Stiles, Mark D},
	journal={Journal of Magnetism and Magnetic Materials},
	volume={320},
	number={7},
	pages={1190--1216},
	year={2008},
	publisher={Elsevier}
}

@article{kawahara2012spin,
	title={Spin-transfer torque RAM technology: Review and prospect},
	author={Kawahara, Takayuki and Ito, Kenchi and Takemura, Riichiro and Ohno, Hideo},
	journal={Microelectronics Reliability},
	volume={52},
	number={4},
	pages={613--627},
	year={2012},
	publisher={Elsevier}
}

@article{li2003magnetization,
	title={Magnetization dynamics with a spin-transfer torque},
	author={Li, Z and Zhang, Shufeng},
	journal={Physical Review B},
	volume={68},
	number={2},
	pages={024404},
	year={2003},
	publisher={APS}
}

@article{barla2021spintronic,
	title={Spintronic devices: a promising alternative to CMOS devices},
	author={Barla, Prashanth and Joshi, Vinod Kumar and Bhat, Somashekara},
	journal={Journal of Computational Electronics},
	volume={20},
	number={2},
	pages={805--837},
	year={2021},
	publisher={Springer}
}

@article{wang2015compact,
	title={Compact thermal modeling of spin transfer torque magnetic tunnel junction},
	author={Wang, You and Cai, Hao and Naviner, Lirida AB and Zhang, Yue and Klein, Jacques-Olivier and Zhao, WS},
	journal={Microelectronics Reliability},
	volume={55},
	number={9-10},
	pages={1649--1653},
	year={2015},
	publisher={Elsevier}
}

@article{emori2013current,
	title={Current-driven dynamics of chiral ferromagnetic domain walls},
	author={Emori, Satoru and Bauer, Uwe and Ahn, Sung-Min and Martinez, Eduardo and Beach, Geoffrey SD},
	journal={Nature materials},
	volume={12},
	number={7},
	pages={611--616},
	year={2013},
	publisher={Nature Publishing Group UK London}
}

@inproceedings{parkin2009racetrack,
	title={Racetrack memory: A storage class memory based on current controlled magnetic domain wall motion},
	author={Parkin, Stuart},
	booktitle={2009 Device Research Conference},
	pages={3--6},
	year={2009},
	organization={IEEE}
}

@article{chen2017skyrmion,
	title={Skyrmion hall effect},
	author={Chen, Gong},
	journal={Nature Physics},
	volume={13},
	number={2},
	pages={112--113},
	year={2017},
	publisher={Nature Publishing Group UK London}
}

@article{luo2021skyrmion,
	title={Skyrmion devices for memory and logic applications},
	author={Luo, Shijiang and You, Long},
	journal={APL Materials},
	volume={9},
	number={5},
	year={2021},
	publisher={AIP Publishing}
}

@article{kezsmarki2015neel,
	title={N{\'e}el-type skyrmion lattice with confined orientation in the polar magnetic semiconductor GaV4S8},
	author={K{\'e}zsm{\'a}rki, Istv{\'a}n and Bord{\'a}cs, S{\'a}ndor and Milde, Peter and Neuber, Erik and Eng, Lukas M and White, Jonathan Stuart and R{\o}nnow, Henrik M and Dewhurst, Charles David and Mochizuki, Masahito and Yanai, K and others},
	journal={Nature materials},
	volume={14},
	number={11},
	pages={1116--1122},
	year={2015},
	publisher={Nature Publishing Group UK London}
}

@inproceedings{yoda2012progress,
	title={Progress of STT-MRAM technology and the effect on normally-off computing systems},
	author={Yoda, H and Fujita, S and Shimomura, N and Kitagawa, E and Abe, K and Nomura, K and Noguchi, H and Ito, J},
	booktitle={2012 International Electron Devices Meeting},
	pages={11--3},
	year={2012},
	organization={IEEE}
}

@article{parkin2008magnetic,
	title={Magnetic domain-wall racetrack memory},
	author={Parkin, Stuart SP and Hayashi, Masamitsu and Thomas, Luc},
	journal={science},
	volume={320},
	number={5873},
	pages={190--194},
	year={2008},
	publisher={American Association for the Advancement of Science}
}

@article{jonietz2010spin,
	title={Spin transfer torques in MnSi at ultralow current densities},
	author={Jonietz, Florian and M{\"u}hlbauer, Sebastain and Pfleiderer, Christian and Neubauer, Andreas and M{\"u}nzer, Wolfgang and Bauer, Andreas and Adams, T and Georgii, Robert and B{\"o}ni, Peter and Duine, Rembert A and others},
	journal={Science},
	volume={330},
	number={6011},
	pages={1648--1651},
	year={2010},
	publisher={American Association for the Advancement of Science}
}

@article{woo2016observation,
	title={Observation of room-temperature magnetic skyrmions and their current-driven dynamics in ultrathin metallic ferromagnets},
	author={Woo, Seonghoon and Litzius, Kai and Kr{\"u}ger, Benjamin and Im, Mi-Young and Caretta, Lucas and Richter, Kornel and Mann, Maxwell and Krone, Andrea and Reeve, Robert M and Weigand, Markus and others},
	journal={Nature materials},
	volume={15},
	number={5},
	pages={501--506},
	year={2016},
	publisher={Nature Publishing Group UK London}
}

@article{yu2016room,
	title={Room-temperature creation and spin--orbit torque manipulation of skyrmions in thin films with engineered asymmetry},
	author={Yu, Guoqiang and Upadhyaya, Pramey and Li, Xiang and Li, Wenyuan and Kim, Se Kwon and Fan, Yabin and Wong, Kin L and Tserkovnyak, Yaroslav and Amiri, Pedram Khalili and Wang, Kang L},
	journal={Nano letters},
	volume={16},
	number={3},
	pages={1981--1988},
	year={2016},
	publisher={ACS Publications}
}

@article{zhang2016magnetic,
	title={Magnetic bilayer-skyrmions without skyrmion Hall effect},
	author={Zhang, Xichao and Zhou, Yan and Ezawa, Motohiko},
	journal={Nature communications},
	volume={7},
	number={1},
	pages={10293},
	year={2016},
	publisher={Nature Publishing Group UK London}
}

@article{slonczewski1989conductance,
	title={Conductance and exchange coupling of two ferromagnets separated by a tunneling barrier},
	author={Slonczewski, John C},
	journal={Physical Review B},
	volume={39},
	number={10},
	pages={6995},
	year={1989},
	publisher={APS}
}

@article{sengupta2016short,
	title={Short-term plasticity and long-term potentiation in magnetic tunnel junctions: Towards volatile synapses},
	author={Sengupta, Abhronil and Roy, Kaushik},
	journal={Physical Review Applied},
	volume={5},
	number={2},
	pages={024012},
	year={2016},
	publisher={APS}
}

@article{sengupta2016magnetic,
	title={Magnetic tunnel junction mimics stochastic cortical spiking neurons},
	author={Sengupta, Abhronil and Panda, Priyadarshini and Wijesinghe, Parami and Kim, Yusung and Roy, Kaushik},
	journal={Scientific reports},
	volume={6},
	number={1},
	pages={30039},
	year={2016},
	publisher={Nature Publishing Group UK London}
}

@article{das2022self,
	title={Self-reset schemes for magnetic domain wall-based neuron},
	author={Das, Debasis and Fong, Xuanyao},
	journal={IEEE Journal on Exploratory Solid-State Computational Devices and Circuits},
	volume={8},
	number={2},
	pages={166--172},
	year={2022},
	publisher={IEEE}
}

@article{das2023bilayer,
	title={Bilayer-skyrmion-based design of neuron and synapse for spiking neural network},
	author={Das, Debasis and Cen, Yunuo and Wang, Jianze and Fong, Xuanyao},
	journal={Physical Review Applied},
	volume={19},
	number={2},
	pages={024063},
	year={2023},
	publisher={APS}
}

@article{brigner2019shape,
	title={Shape-based magnetic domain wall drift for an artificial spintronic leaky integrate-and-fire neuron},
	author={Brigner, Wesley H and Hassan, Naimul and Jiang-Wei, Lucian and Hu, Xuan and Saha, Diptish and Bennett, Christopher H and Marinella, Matthew J and Incorvia, Jean Anne C and Garcia-Sanchez, Felipe and Friedman, Joseph S},
	journal={IEEE Transactions on Electron Devices},
	volume={66},
	number={11},
	pages={4970--4975},
	year={2019},
	publisher={IEEE}
}

@inproceedings{das2022design,
	title={Design of Spintronics-based Neuronal and Synaptic Devices for Spiking Neural Network Circuits},
	author={Das, Debasis and Cen, Yunuo and Wang, Jianze and Fong, Xuanyao},
	booktitle={2022 IEEE International Conference on Emerging Electronics (ICEE)},
	pages={1--6},
	year={2022},
	organization={IEEE}
}

@article{sengupta2016hybrid,
	title={Hybrid spintronic-CMOS spiking neural network with on-chip learning: Devices, circuits, and systems},
	author={Sengupta, Abhronil and Banerjee, Aparajita and Roy, Kaushik},
	journal={Physical Review Applied},
	volume={6},
	number={6},
	pages={064003},
	year={2016},
	publisher={APS}
}

@article{yen2019all,
	title={All nonmetal resistive random access memory},
	author={Yen, Te Jui and Gismatulin, Andrei and Volodin, Vladimir and Gritsenko, Vladimir and Chin, Albert},
	journal={Scientific reports},
	volume={9},
	number={1},
	pages={6144},
	year={2019},
	publisher={Nature Publishing Group UK London}
}
\end{document}